\documentclass[journal=mamobx,layout=twocolumn]{achemso}

\usepackage{graphicx}
\usepackage{amsmath,amssymb,latexsym}

\usepackage[version=3]{mhchem} 

\newcommand{\mean}[1]{{\left< #1 \right>}}

\author{M. Baiesi}
\email{baiesi@pd.infn.it}
\affiliation[University of Padua]{Department of Physics and Astronomy, University of Padua, Via Marzolo 8, I-35131 Padova, Italy}
\alsoaffiliation[INFN, Sezione di Padova]{INFN, Sezione di Padova, Via Marzolo 8, I-35131 Padova, Italy}
\author{E. Orlandini}
\affiliation[University of Padua]{Department of Physics and Astronomy, University of Padua, Via Marzolo 8, I-35131 Padova, Italy}
\alsoaffiliation[INFN, Sezione di Padova]{INFN, Sezione di Padova, Via Marzolo 8, I-35131 Padova, Italy}
\author{A. L. Stella}
\affiliation[University of Padua]{Department of Physics and Astronomy, University of Padua, Via Marzolo 8, I-35131 Padova, Italy}
\alsoaffiliation[INFN, Sezione di Padova]{INFN, Sezione di Padova, Via Marzolo 8, I-35131 Padova, Italy}

\title{Knotted globular ring polymers: how topology affects statistics and thermodynamics}

\begin{document}


\begin{abstract}
  The statistical mechanics of a long knotted collapsed polymer is determined by a free-energy
with a knot-dependent subleading term, which is linked to the length of the shortest
polymer that can hold such knot. The only other parameter depending on the knot kind
is an amplitude such that relative probabilities of knots do not vary with the temperature $T$,
in the limit of long chains. We arrive at this conclusion by simulating 
interacting self-avoiding rings at low $T$ on the cubic lattice, both
with unrestricted topology and with setups where the globule is divided 
by a slip link in two loops  (preserving their topology)
which compete for the chain length, either in contact or separated
by a wall as for translocation through a membrane pore.
These findings suggest that in macromolecular environments
there may be entropic forces with a purely topological origin, whence
portions of polymers holding complex knots should tend to expand at the expense
of significantly shrinking other topologically simpler portions.
\footnote{This document is the unedited Author's version of a Submitted Work that was subsequently accepted for publication in Macromolecules, copyright \copyright American Chemical Society after peer review. To access the final edited and published work see DOI 10.1021/ma5020287}
\end{abstract}


\section{Introduction}
A main goal of topological polymer statistics is that of determining
up to what extent the thermodynamics of macromolecular chains is affected by
the presence of a fixed topological entanglement~\cite{Orlandini&Whittington:2007:Rev-Mod_Phys,Deguchi-tsurusaki:1997,IdealKnotsBook, Dai:2001,Renner:2014,Matthews:2010:EPL,Matthews:2012:Macrolett,Narros:2013:Macro,Poier:2014:Macro,Katritch:1996:Nature,Arsuaga:2005:Proc-Natl-Acad-Sci-U-S-A:15958528,Marenduzzo:2009:Proc-Natl-Acad-Sci-U-S-A:20018693,Tang:2011:PNAS,Tubiana:2011:PRL,Marenduzzo:2013:PNAS,Katritch:2000:PRE,Farago:2002:EPL,Marcone:2005:J-Phys-A,Marcone:2007:PRE,Rawdon:Macromol:2008a,Rawdon:Macromol:2008b,Orlandini:2003:PRE,Hanke:2003:PRE,Orlandini:2004:J-Stat-Phys,Baiesi:2011:PRL,Baiesi:2007:PRL}.
Clarifying issues related to knots and links in polymer physics has been since long 
recognized of key importance in a variety of fields, ranging from molecular biology 
to nanotechnology~\cite{Arai:1999:Nature,Ayme:2012:NC,Podtelezhnikov:1999:PNAS,Arsuaga:2005:Proc-Natl-Acad-Sci-U-S-A:15958528,Marenduzzo:2009:Proc-Natl-Acad-Sci-U-S-A:20018693,Tang:2011:PNAS,Tubiana:2011:PRL,Marenduzzo:2013:PNAS,Tkalec:2011,Irvine:2014}.
So far, for isolated ring polymers, which are the object of the present study, most 
theoretical and numerical approaches provided
informations referring to ensembles with unrestricted topology. This means
that information on the frequency of occurrence of different knots is in general
not available for models of random ring polymer configurations. One
does not have an idea of the extent to which, at least for long chains, these
frequencies could reveal universal features, possibly connected to topological
invariants. Thus, it remains unclear how different topologies
contribute to the global free energy of systems described in such ensembles.
This situation contrasts with the fact that ring polymers are not expected to change their 
topology in most experimental situations. Another
fundamental problem is that of understanding if the presence of a fixed
topology can give rise to thermodynamic effects, which would not
be revealed within mixed topology ensembles.

The difficulty of answering questions like those above and the complexity of
the scenarios the answers outline, depend in first place on the degree of
localization of topological entanglement in the conditions considered
for the polymer~\cite{Katritch:2000:PRE}.
For example, when dealing with polymers in good solvent, we
know that the prime components of a knot are weakly
localized~\cite{Farago:2002:EPL,Marcone:2005:J-Phys-A,Marcone:2007:PRE}.
Each prime component of the knot behaves almost like a small bead sliding along the backbone.
This makes it relatively simple to guess the gross structure of finite
size corrections to the free energy expected for a knotted ring~\cite{Baiesi:2010:JSM}. The form
of these corrections and the behavior of asymptotic relative frequencies of different
knots are now known in great detail thanks to studies based on samples of configurations
of extremely long knotted self-avoiding rings on a
lattice~\cite{Orlandini:1998:J-Phys-A,Yao:2001:J-Phys-A,Baiesi:2010:JSM,Rensburg&Rechnitzer:2011:JPA,Baiesi:2012:PRE}.
Indeed, while as established also by rigorous theorems~\cite{Sumners&Whittington:1988:J-Phys-A,Diao:1994:JKTR}, unknotted configurations in good solvent
should occur with zero frequency in infinite chains, it is necessary to consider
very long rings in order to observe with appreciable frequency knotted configurations.
For swollen polymers the decay to zero with increasing backbone length of the frequency of initially
dominant, unknotted configurations is too slow to allow a
rich enough sampling of knots with short rings.

The situation is quite different, and in several aspects more interesting,
in the case of globular ring polymers. For chains in bad solvent
it is well established that topological entanglement is delocalized and spreads
along the whole backbone~\cite{Marcone:2005:J-Phys-A,Orlandini:2003:PRE,Hanke:2003:PRE,Orlandini:2004:J-Stat-Phys,Baiesi:2011:PRL}.
This delocalization is accompanied by the fact that
models for generating random polymer configurations show a rich spectrum of
knots already for relatively short chains. The unknot in this case does not
occur with zero probability relative to any knotted configuration in the
limit of infinitely long rings. To the contrary, there is evidence of an
asymptotic spectrum in which the relative frequencies of all knots with respect to
the unknot approach finite, nonzero limits~\cite{Baiesi:2007:PRL}. Additional interest
in the globuli is due to some recently discovered~\cite{Baiesi:2011:PRL}, remarkable
thermodynamic properties directly linked to topology, which manifest themselves
in processes like translocation through membrane pores.

\begin{figure}[tb]
\includegraphics[width=0.8\columnwidth]{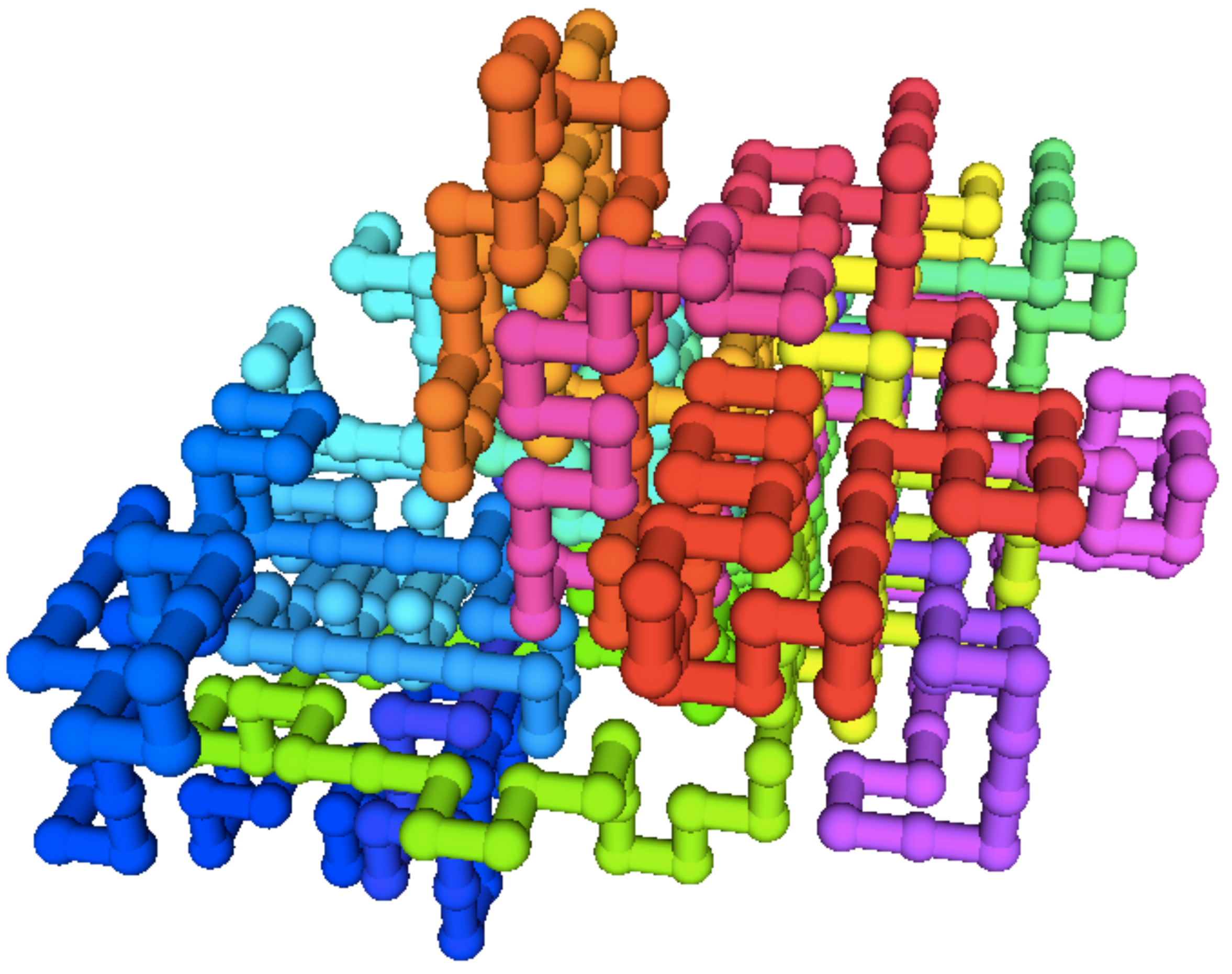}\\
\includegraphics[width=0.57\columnwidth]{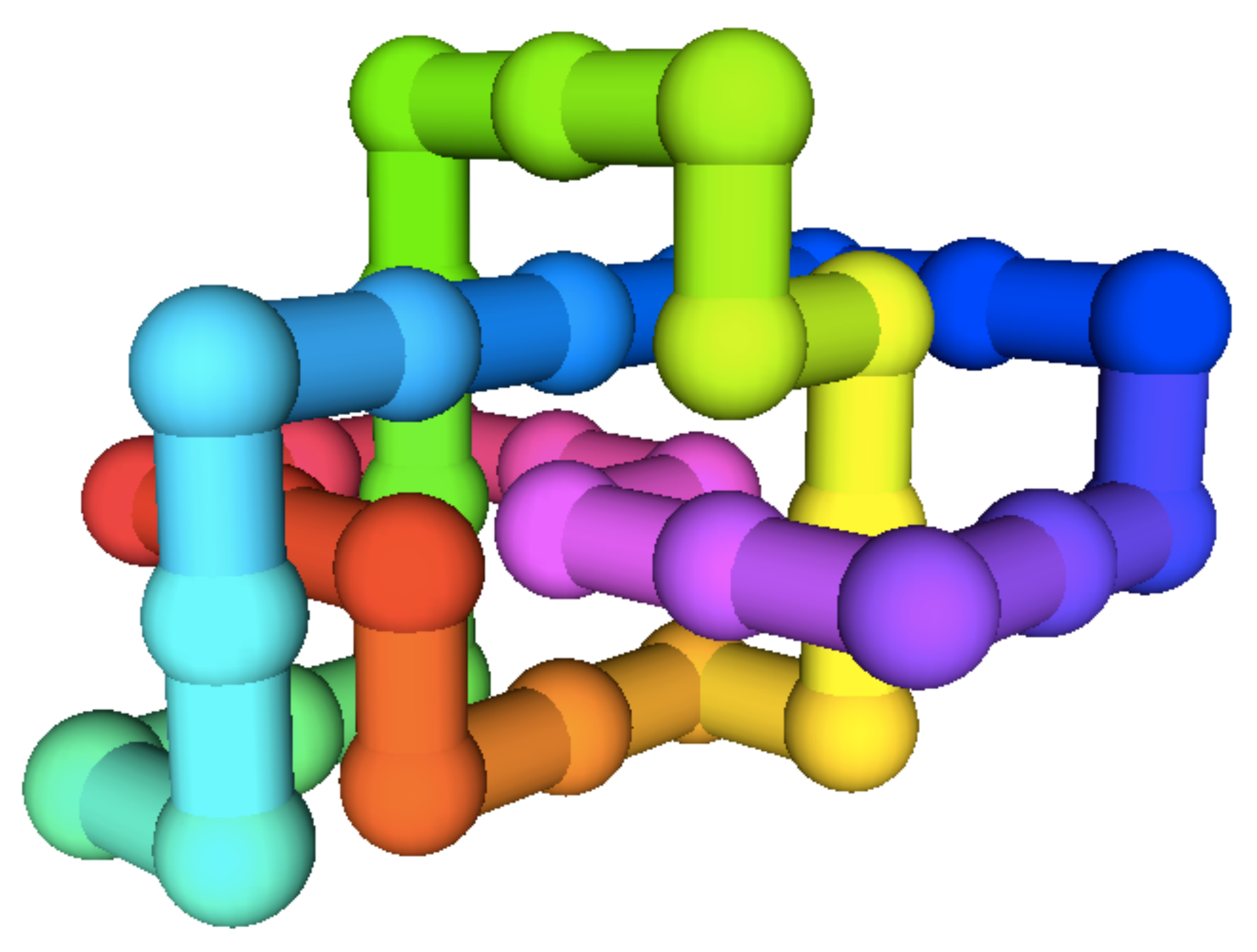}
\includegraphics[width=0.37\columnwidth]{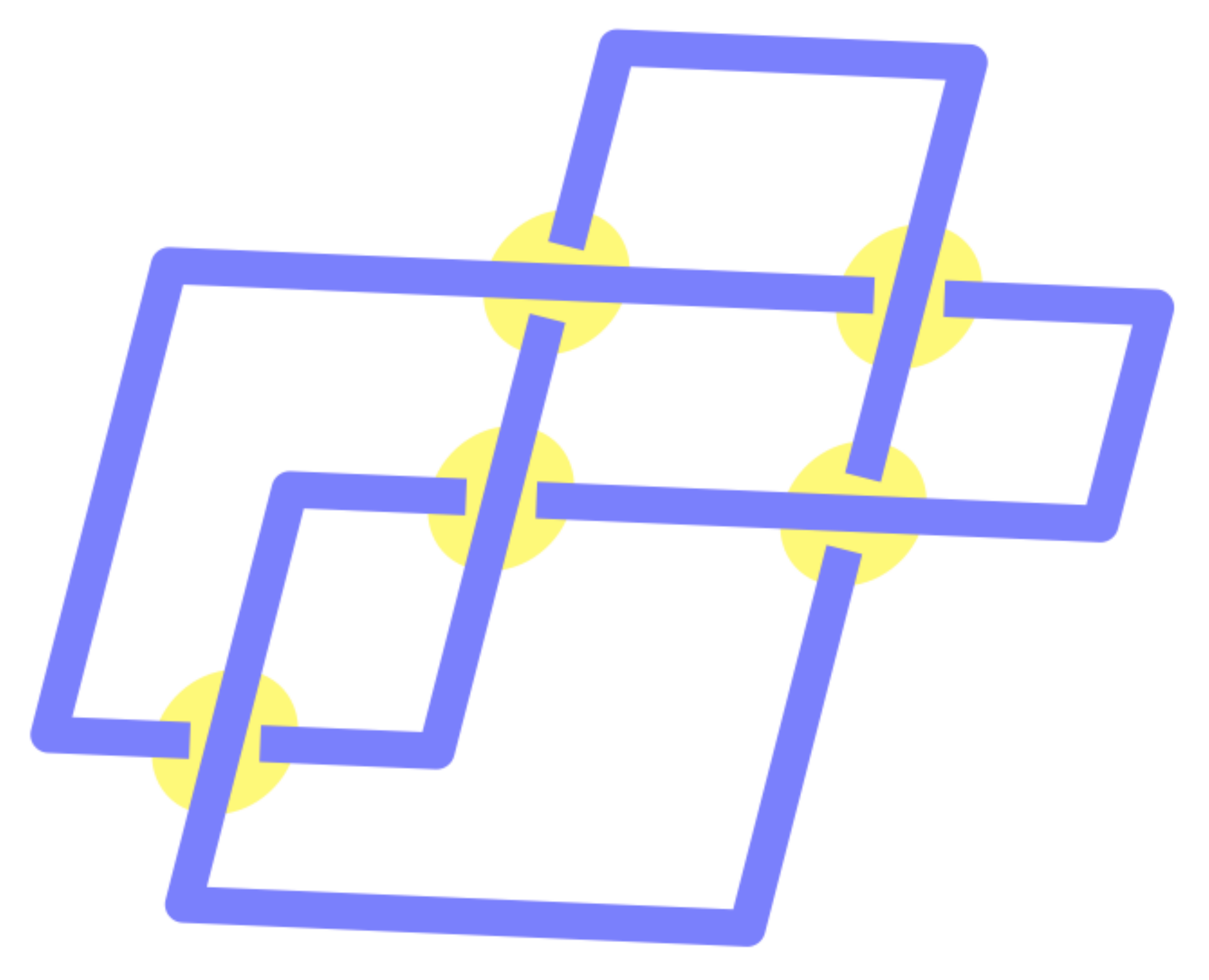}
\caption{Example of a collapsed interacting self-avoiding ring with $N=500$ steps on the cubic lattice,
and its shortened form after the BFACF~\cite{Aragao&Caracciolo:1983:J-Physique,Berg&Foester:1981:Phys-Lett-B} 
reduction procedure with a low fugacity per step.
The latter reveals that the ring holds a $5_2$ knot, 
whose projection with the minimal crossing number $n_c=5$ is also shown.
}
\label{fig:BFACF}
\end{figure}

The first steps towards a characterization of the spectrum of knots in ring polymers
in the globular phase were made in Ref.~\cite{Baiesi:2007:PRL}. Due to the difficulty of
sampling dense configurations by Monte Carlo, the investigation was limited to a
single temperature below the $\Theta$ point of self-avoiding rings on cubic lattice
with attractive nearest neighbor interactions. This study showed that the relative frequencies
of different knots tend to finite asymptotic values and have a ranking consistent
with a Zipf law. A further result concerns the dependence of the globular free
energy on the topological invariants of the knot: while in the infinite ring limit
the free energy per monomer is independent of topology, a dependence on
the minimal crossing number ($n_c$) of the knot enters in one of its subleading finite
size corrections Ref.~\cite{Baiesi:2007:PRL,Baiesi:2011:PRL} 
(an example of knot with  $n_c=5$ is shown in Fig.~\ref{fig:BFACF}). 
This last result suggests that the value of $n_c$ 
could play a relevant role in the thermodynamics of a knotted globule.
A way to shed light on this role consists in studying the behavior
of the globule under geometrical constraints which interfere with the
topology. In Ref.~\cite{Baiesi:2011:PRL} this kind of investigation was first performed
using slipping links which divide the single globule in two interacting and
possibly knotted loops in equilibrium. The simulations showed that indeed the
respective minimal number of crossings precisely determines the way in which
the two loops share the total ring length on average. The findings of
Ref.~\cite{Baiesi:2011:PRL} on the topological correction to the free energy per monomer
were also the basis for explaining a remarkable and unexpected phenomenon
occurring when the slip link is replaced by a hole in a repulsive plane
separating the ring in two non-interacting globuli with different topologies.
This configuration schematizes a ring polymer translocating through a membrane
pore. The different minimal number of crossings of the two knotted loops induces an
asymmetry in the free energies of the competing globuli favoring configurations
in which the globule with more crossings keeps most of the available
backbone length. This asymmetry thus distorts the two otherwise symmetrical 
global free energy minima due to the surface tension of the globuli.

While being of fundamental interest, the problems mentioned above
should be directly relevant for practical applications. Besides the 
translocation processes, of central importance for biology,
the study of topological spectra of DNA extracted from viral capsids
is a typical context where the knowledge of spectra for specific models
of random polymers is very 
valuable~\cite{Arsuaga:2005:Proc-Natl-Acad-Sci-U-S-A:15958528,Marenduzzo:2009:Proc-Natl-Acad-Sci-U-S-A:20018693,Tubiana:2011:PRL}.
Model calculations like those carried
out in the present work could also help in addressing issues like the known
relative rarity of knots one finds in native globular proteins~\cite{Virnau:2006:PLoS-Comput-Biol:16978047,Sulkowska12162008,Potestio:2010:PLoS-Comput-Biol:20686683}.
The investigation of the effects of topological constraints on globular
polymers, both in and out of equilibrium, is also expected to be a
key to understand chromosomal architecture~\cite{Dorier:2009:Nucleic-Acid-Res,Grosberg:2012:PolScie}.

In the present work we considerably extend and discuss in further detail some of 
the results presented in Refs.~\cite{Baiesi:2007:PRL,Baiesi:2011:PRL}.
Rather than assuming the more global perspective implicit in
the Zipf law, here we look at the relative asymptotic frequencies of the simplest
knots in globular ring polymers. The results furnish a clear evidence of the independence
of temperature of these frequencies in the whole region below the $\Theta$ point and
suggest the intriguing possibility that they represent universal, topology related
numbers. This investigation proceeds in parallel with a better
determination of the topological correction to the free energy of the globuli
already postulated in~\cite{Baiesi:2011:PRL}. The analysis of data in a whole range of low
temperatures gives stronger support to the conjectured form and corroborates
the explanation of the already mentioned bias due to topology in the translocation
of the globuli. Our analysis also shows that the strength of the topological
correction is connected to the length of the knots in their ideal, minimal
length form~\cite{Katritch:1996:Nature,IdealKnotsBook,Rensburg&Promislow:1995:JKTR,Diao:1993:JKTR,Janse:2011:JSM}. 
This connection sheds light on
the reason why the correction itself seems to be determined primarily and
almost exclusively by the minimal crossing number of the knots. 

This article is organized as follows. In the next section we define the
model and describe the methods of Monte Carlo simulation used for our
analysis. Section {\em Relative frequencies and free energies of knotted globuli} 
is devoted to the analysis of the data for
relative frequencies at different temperatures and to their extrapolation
for infinitely long rings. The form of the finite size topological
correction to the free energy and its dependence on both the minimal number of
crossings and temperature are discussed in this section. In Section 
{\em The role of $n_c$ in the thermodynamics of the globule}
we present results concerning the interference of geometrical constraints
due to slip-links with topology.
In Section {\em Effects of the topological correction in translocation} 
we refine our discussion of the effects of constraints
due to translocation set-ups, and verify their consistency with the
postulated topological finite size correction for the globular free energy.
We also discuss the  evidence of a connection between the topological correction 
and the properties of ideal knots.
The last section is devoted to conclusions.

\section{Model and Simulations}

We model cyclic, flexible polymers with excluded volume as
$N$-step self-avoiding rings on a cubic lattice (see Fig.~\ref{fig:BFACF}). In order to induce collapse
at low enough temperature $T$, we introduce an attractive interaction
($\epsilon =-1$) between nearest neighbors sites visited by the
ring which are not consecutive along the backbone.

\subsection{Variable topology}
A first problem encountered in our analysis is that of generating
a sufficient number of uncorrelated collapsed configurations  by Monte Carlo.
To this purpose we have to fix the temperature below the $\Theta$ point, which in our units
is at $T_{\Theta} \simeq 3.72$~~\cite{Tesi_et_al:1996:J_Phys_A,Grassberger:1997:PRE}.
The pruned enriched Rosenbluth method (PERM) is efficient in generating
collapsed polymer configurations~\cite{Grassberger:PRE:2002,Baiesi:2007:PRL}; as discussed below,
we could obtain reasonably rich samplings
down to $T = 1.92$, for chains with length up to $N=1400$.

Of course, the study of the topological spectrum of
the generated ring configurations involves another task, which is even more demanding:
that of determining the knot type of each sampled configuration.
This constitutes the bottleneck of our simulations.
Indeed, it turns out to be a relatively minor problem the fact that PERM generates open chains and that we
must discard most of them, keeping only those becoming rings upon addition of a further step.

Since the collapsed configurations are geometrically very intricated,
with planar projections presenting huge numbers of crossings,
before attempting a successful analysis of the knot type we need
to simplify each configuration while keeping its topology
unaltered. This simplification is achieved with
a grand-canonical algorithm of the BFACF 
type~\cite{Madras&Slade:1993,Aragao&Caracciolo:1983:J-Physique,Berg&Foester:1981:Phys-Lett-B},
which has the property of preserving the knot type
because it lets the configurations evolve only with local and
crankshaft deletion/insertion moves. In our case the simplification is
achieved with a bias toward deletion that emerges from choosing
a low fugacity $K$ per step. After a rapid shrinking of the ring (Fig.~\ref{fig:BFACF}),
it is eventually possible to analyze the knot type on the basis of
HOMFLY polynomials within the "Knotscape" program~\cite{knotscape}.

\subsection{Fixed topology}
For studying the effects of external geometrical constraints on globular rings
with fixed topology, we apply the BFACF grand canonical simulation
method.
As already explained in Ref.~\cite{Baiesi:2011:PRL}, there is
a drawback of this method when applied to polymers below the $\Theta$ point:
The length $N$ of the generated configurations does not grow continuously to
$+ \infty$ with the step fugacity, upon approaching its critical value
from below~\cite{Baiesi:2011:PRL}. This behavior, consistent with the tricritical character
of the $\Theta$ point, is due to the fact that the globule
for $T<T_{\Theta}$ has an interfacial free energy growing
as its surface, i.e. $\sim N^{2/3}$, in addition to the bulk
one growing like $N$. 
As a consequence, it is not possible to gradually tune the length of the simulated rings
just by varying $K$. In order to do so, as already explained in
Ref.~\cite{Baiesi:2011:PRL}, we introduce an extra $N$-dependent Gaussian weight
multiplying the grand-canonical one for the configurations (an alternative method
was recently proposed in~\cite{Baiesi:2014:PRE}). This
weight is maximum at a tunable $\overline{N}$ and has a sufficiently small
width, so that $N\sim \overline{N}$ during the simulation. Thus, the sampling is essentially
canonical up to relatively small fluctuations in $N$.
This canonical character is needed in the analysis of
how the competition between geometry and topology evolves
as a function of the length of the rings, keeping the other
parameters fixed.

\section{Relative frequencies and free energies of knotted globuli}

The form of the free energy of a ring polymer at $T<T_{\Theta}$
can be guessed on the basis of an analogy with the swollen case and
of the peculiar physical features of the globular phase.
In particular, we can expect two facts: i) the free energy per monomer $f$
in the bulk of the globules is lower than the free energy $f_s$ of monomers
on the interface between the globule and the solvent; ii) the number of monomers
on the surface of a (smooth) globule scales asymptotically as $N^{2/3}$.
Thus, a plausible
starting ansatz~\cite{Owczarek_PB_PRL93:collapsed,Baiesi:2007:PRL}
for the canonical partition function of a collapsed
ring with $N$ steps, in an ensemble with unrestricted topology, is
\begin{equation}
Z_N(T) \simeq A e^{\kappa N} e^{\sigma N^{2/3}} N^{\alpha-2}
\end{equation}
where $\kappa =-\frac{f}{k_B T} >0$
and $\sigma N^{2/3}$ is a dimensionless interfacial correction to the bulk free energy.
Indeed the number of monomers on the smooth globular surface
grow as $N^{2/3}$, and $\sigma < 0$.
The factor $N^{\alpha -2}$ is just written in
analogy with the swollen case and accounts for a possible further non
extensive logarithmic correction to the free energy.
In fact even tentative estimates of $\alpha$ for the globular phase
were not available until quite recently, when some of the present authors
proposed a Monte Carlo method for its determination~\cite{Baiesi:2014:PRE}.
Here we focus on an ansatz for $Z_{k,N}$, the partition function of a
globule with specific knot $k$. As a first hypothesis we could expect that
at least some of the quantities in Eq.~(1) assume a dependence on $k$ in
order to represent the physics of a globule with fixed knot:
\begin{equation}
\label{Zk}
Z_{k,N} \simeq A_k e^{\kappa_k N} e^{\sigma_k N^{2/3}} N^{\alpha_k-2} 
\end{equation}
Studies of the self-avoiding rings at $T=\infty$, for which $\sigma=0$,
have shown that
in the swollen phase the amplitude $A_k$ indeed depends on $k$, while
$\kappa_k$ does not seem to vary with the knot type~\cite{Orlandini:1998:J-Phys-A,Baiesi:2010:JSM}
For all temperatures $T> T_{\Theta}$  is also expected
(and known exactly in the limit $T\to \infty$~\cite{Sumners&Whittington:1988:J-Phys-A})
that $\kappa_{\emptyset} < \kappa$ where $k=\emptyset$ denotes the unknot.
Remarkable in that case is the marked $k$ dependence of $\alpha$, which
is increasing by one unit for each prime component of
$k$~\cite{Orlandini:1998:J-Phys-A,Baiesi:2010:JSM}.
Thus, the frequencies of different knots relative to
the unknot turn out to diverge for $N \to \infty$ with a factor
$N$ raised to a power equal to the number of prime components
of $k$. This behavior as a function of $N$ is a strong
confirmation that $\kappa_k$ does not depend on $k$ in that case.

\subsection{Knot frequencies}
We have sampled collapsed rings with the PERM algorithm at five
different temperatures well below the $\Theta$ point:
$1/T = 0.37$, $0.4$, $0.44$, $0.48$, $0.52$. The $N$'s chosen for sampling
are $400$, $500$, $600$, $700$, $800$, $1000$, $1200$ and
$1400$. However, in some cases we will show only data up to $N=1000$ or $N=1200$
because the statistics (especially of some complex knots) was not rich enough
for the larger $N$'s at the lower temperatures.

\begin{figure}[tb]
\includegraphics[width=0.95\columnwidth]{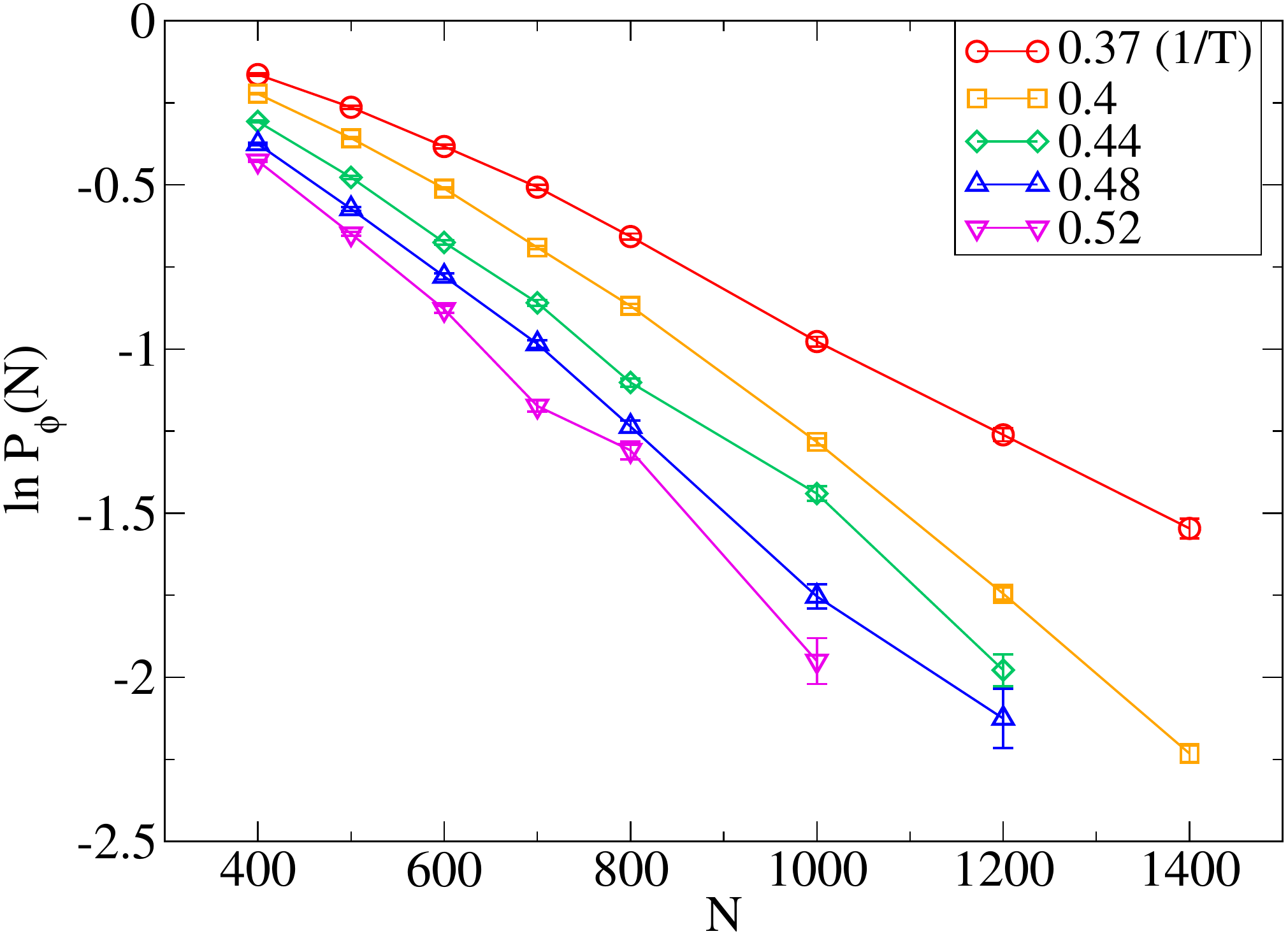}
\caption{Logarithm of the frequency of unknot configurations as a function of $N$, for five different temperatures.
}
\label{fig:P0}
\end{figure}

When analyzing the globular phase, a first key property to study
is the (absolute) probability $P_\emptyset$ of the unknotted configuration
within the ensemble with unrestricted topology. As in the case of
$T \to \infty$, the frequency of unknotted configurations is
exponentially decreasing with $N$, but much more rapidly. As
shown in Fig.~\ref{fig:P0}, the decay constant
$N_\emptyset$ is of the order of $400$ for all $T$, while for self-avoiding polygons
without attraction it would be of the order of $200000$. This
explains why simulations in the globular phase generate a much
richer variety of knotted configurations than in the swollen phase
already for moderately long chains.
In Table~\ref{tab:1} we list the estimated $N_\emptyset$ values for all temperatures.
A trend of $N_\emptyset$ becoming lower with the temperature is evident.
For $T\to 0$ one might expect that $N_\emptyset$ tends to a value similar or equal to
the one estimated for Hamiltonian walks~\cite{LUA:2004:Pol}. 

\begin{table}[!b]
\begin{center}
\begin{tabular}{ c  c  c c  c }
\hline
$1 / T$\ \ & $N_\emptyset$ & $C(T)$ & $a$ &  $C(T)$ \\
\hline
0.37 & 678(14) & 242 & 1.30 & 223 \\
0.40 & 450(30) & 198 & 1.33 & 197\\
0.44 & 460(45) & 171 & 1.33 & 170\\
0.48 & 435(25) & 154 & 1.33 & 155\\
0.52 & 380(40) & 138 & 1.36 & 145\\
\hline
\end{tabular}
\end{center}
\caption{Some parameters for data at different temperatures: the
$N_\emptyset$ for the exponential decay of the probability of unknots, the
$C(T)$ \& $a$ for fits $C(T) n_c^{a}$ of data in Fig.~\ref{fig:Mk}, and  $C(T)$ of the fits
$C(T) n_c^{1.33}$ shown in Fig.~\ref{fig:Mk}.
}
\label{tab:1}
\end{table}

\begin{figure}[tb]
\includegraphics[width=0.95\columnwidth]{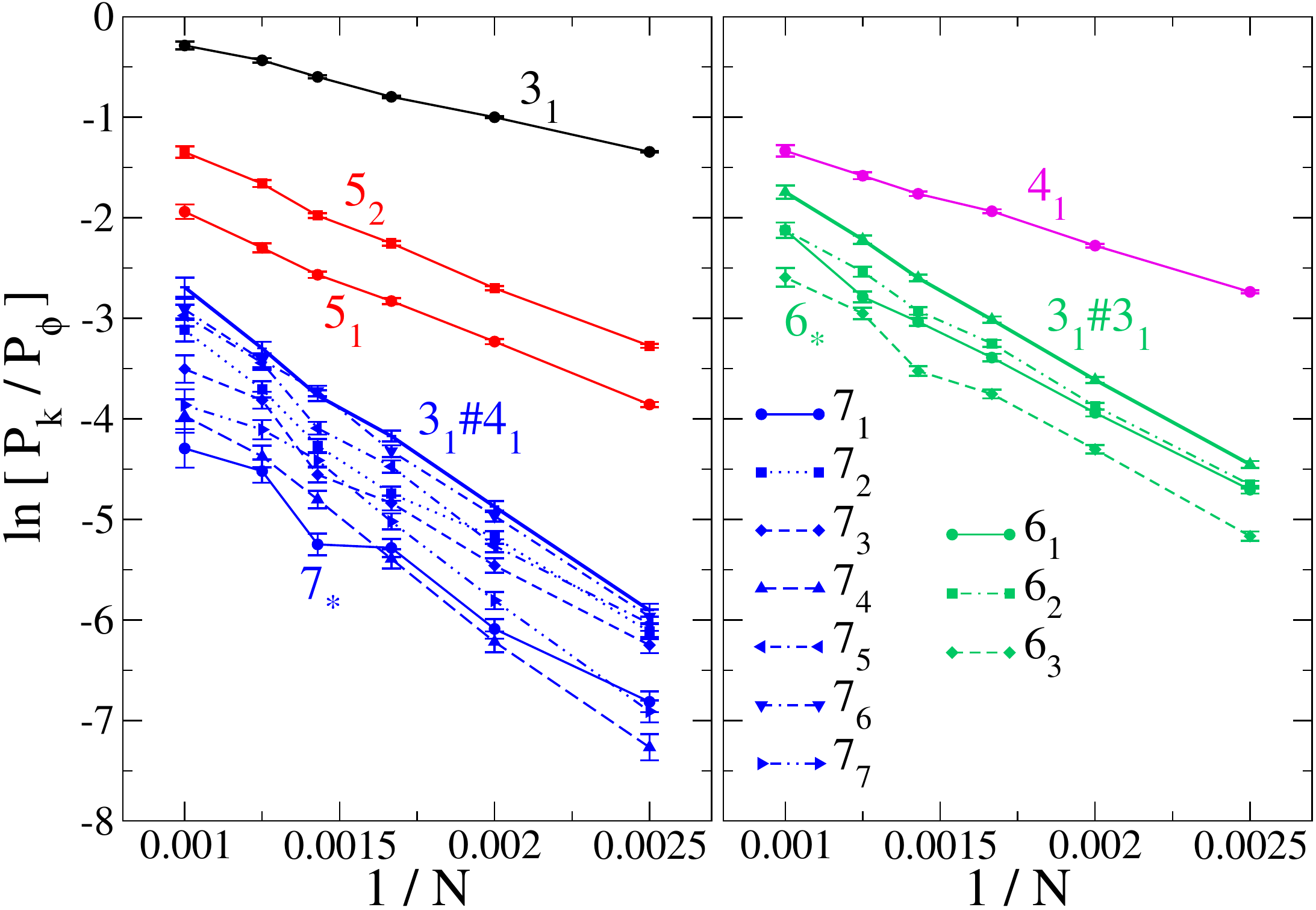}
\caption{Frequency of knot types relative to the frequency of the unknot,
 at temperature $T=1/0.48$ in the collapsed phase, as a function of $1/N$.
Colors distinguish knots according to their minimal crossing number. There are also two cases
of composite knots (thick lines).
}
\label{fig:1N}
\end{figure}

We made a systematic analysis of the frequencies $P_k$ of knots
of increasing complexity, relative to that of the unknot. A first summary of
our findings is provided in Fig.~\ref{fig:1N} where we plot the relative
frequencies at $1/T=0.48$ for the first few knots
and for chain length ranging from $N=400$ up to $N=1000$.
The plots report the logarithms of the frequencies as a function
of $\frac{1}{N}$, and show two remarkable features: in all cases
the extrapolation to $N\to \infty$ appears to give a finite limit;
furthermore, the linearity of the plots suggests a correction
$\sim 1/N$ for the total free energy of the globules. The slope
is approximately the same for different knots having the same
minimal number of crossings, $n_c$. From the fact that the relative
frequencies approach finite limits, we argue that, at the given
temperature, if Eq.~(2)	applies, the parameters $\kappa_k$, $\sigma_{k}$
and $\alpha_k$ should be independent of $k$, while only
the amplitudes $A_k$ should account for the different asymptotic
frequencies. This could have been expected for the first two parameters,
but marks an important difference from the swollen case as far as $\alpha_k$ is
concerned. On the other hand, the $k$ dependence of $\alpha_k$ for
$T\to \infty$ clearly originates from the localized character of prime
knots, which does not hold in the globular phase.

Fig.~\ref{fig:1N} also suggests
that there is a finite size correction to the free energy whose topology
dependence is, at least to a good approximation, only through the minimal
number of crossings of the knot $n_c$. The relative character of our
frequency determinations does not in principle exclude that
the part of globule free energy independent of topology
could have a form different from that implied by
Eq.~(3) with $k$ independent parameters. In particular other topology
independent finite size corrections could be present.
In this work we focus on the
topology dependent correction and on its consequences for the physics of
the globule.

\begin{figure*}[tb]
\includegraphics[width=0.65\columnwidth]{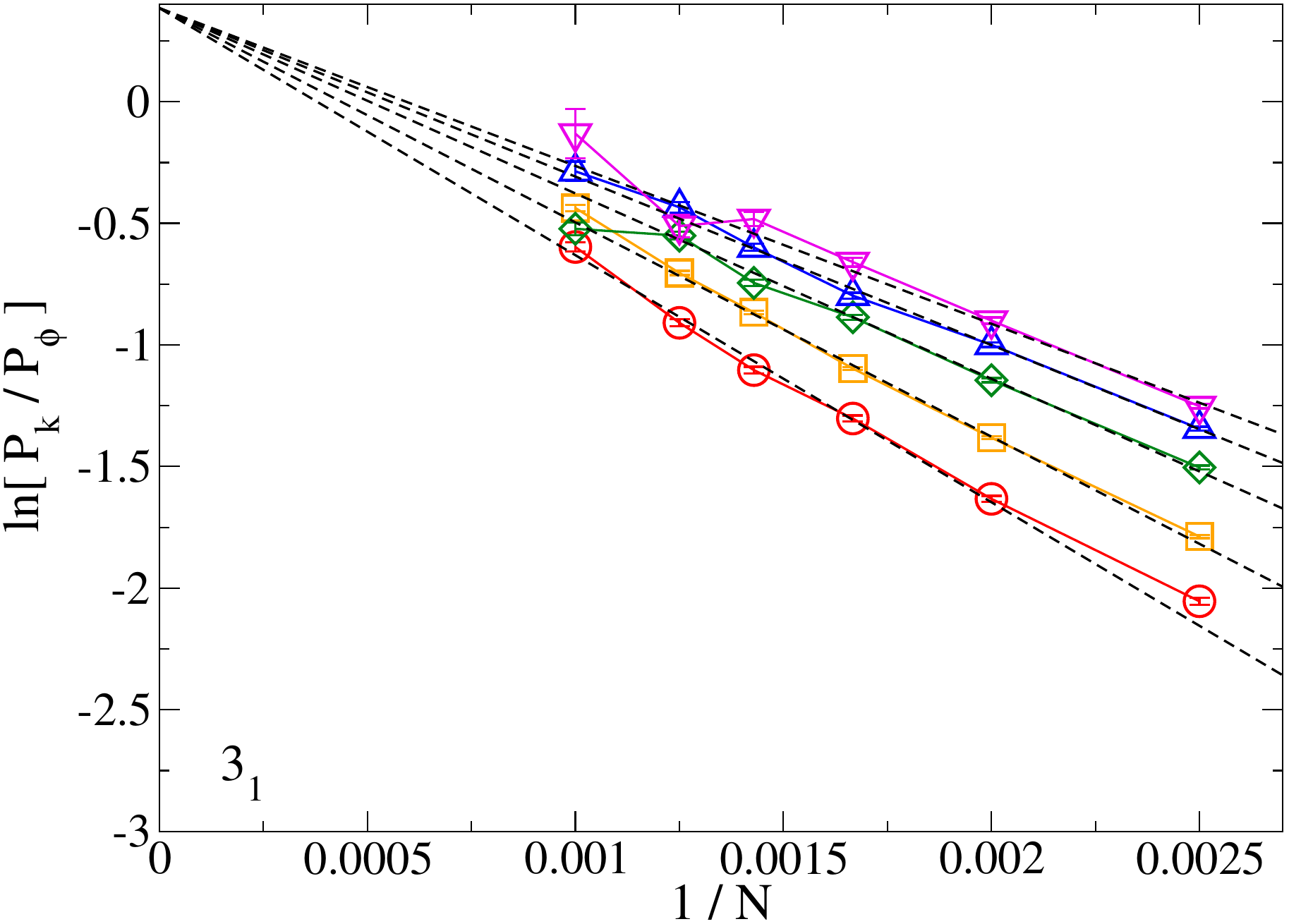}
\hskip 0.05\columnwidth
\includegraphics[width=0.65\columnwidth]{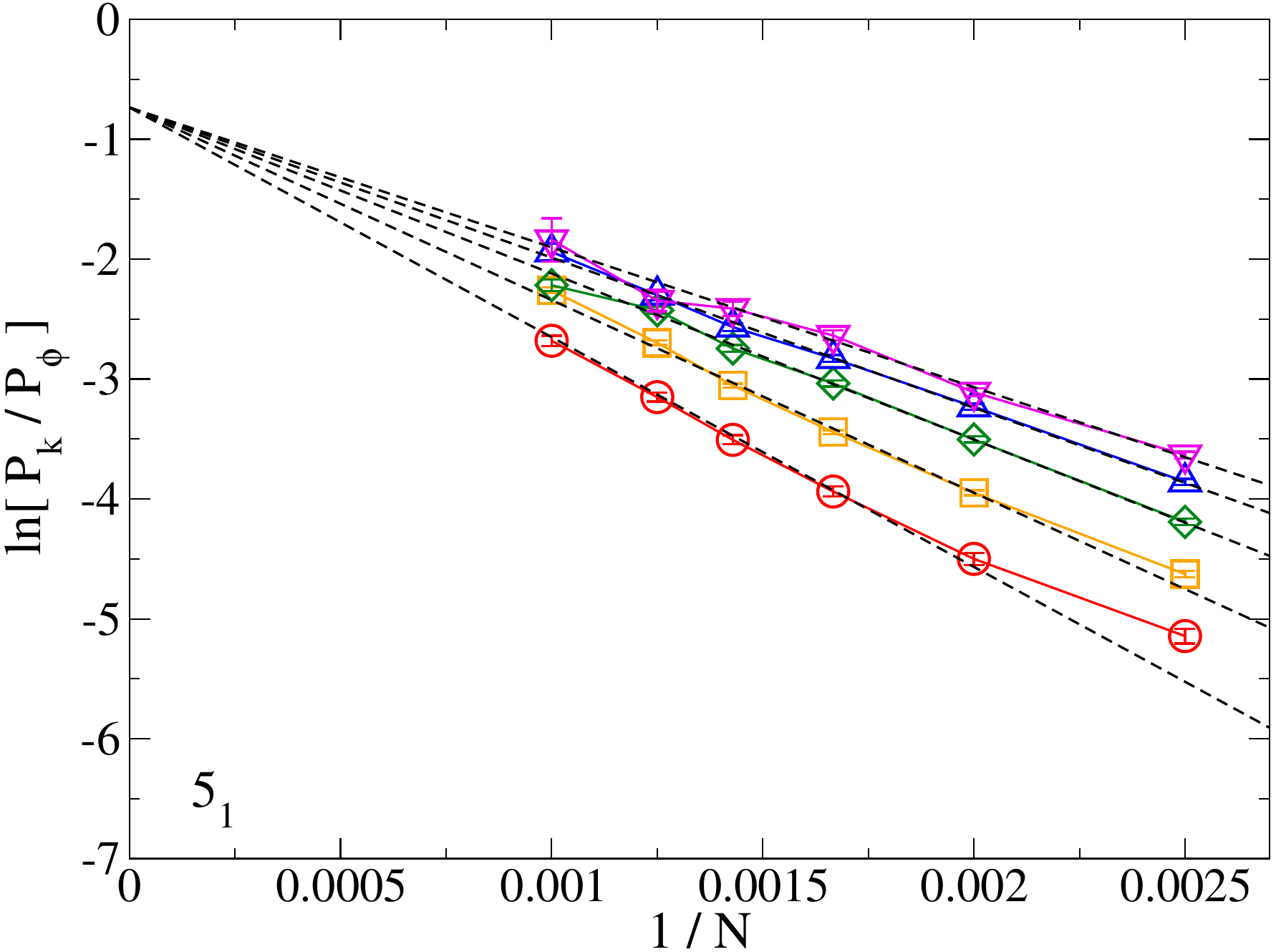}
\hskip 0.05\columnwidth
\includegraphics[width=0.65\columnwidth]{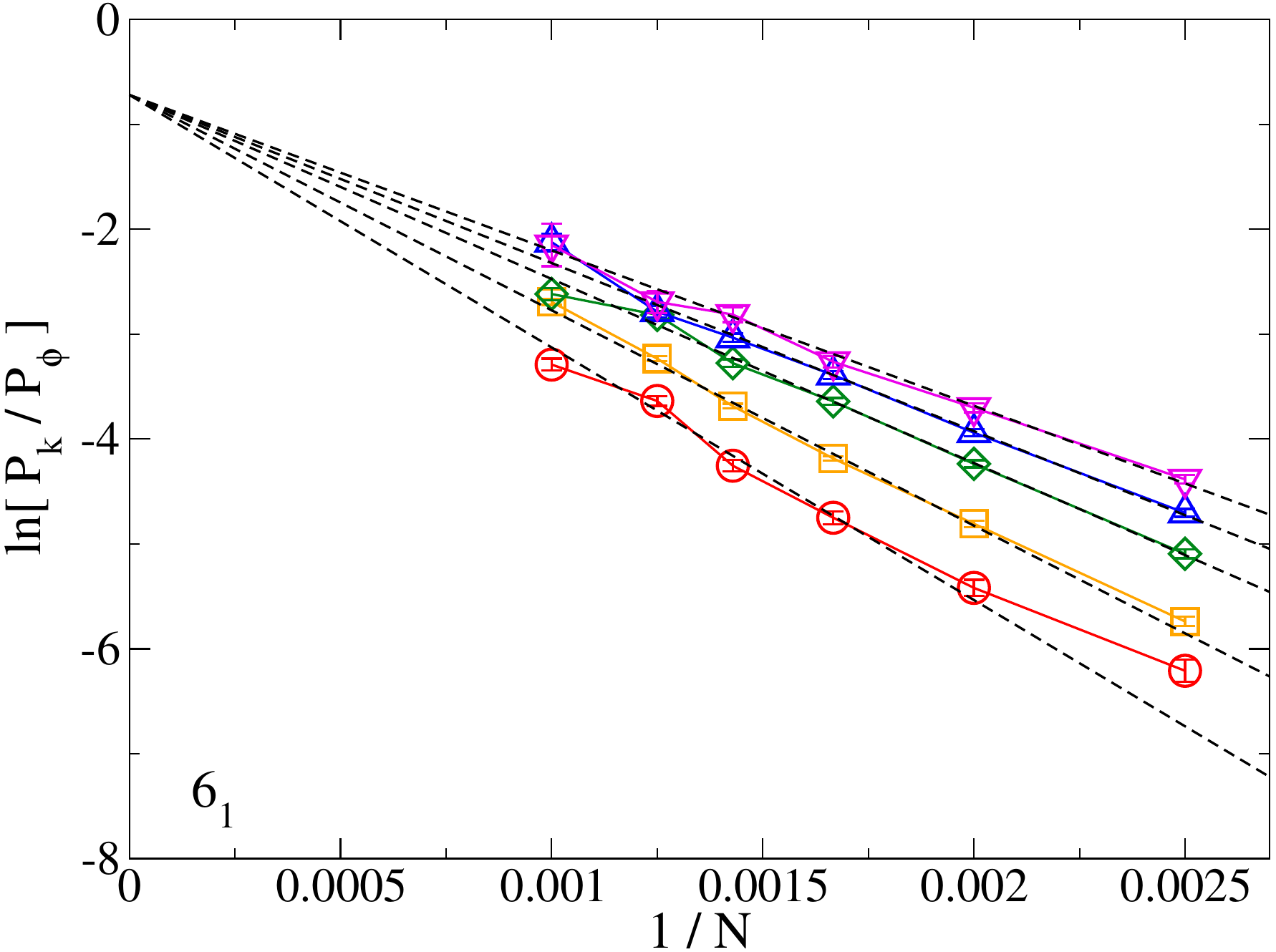}
\vskip 0.05\columnwidth
\includegraphics[width=0.65\columnwidth]{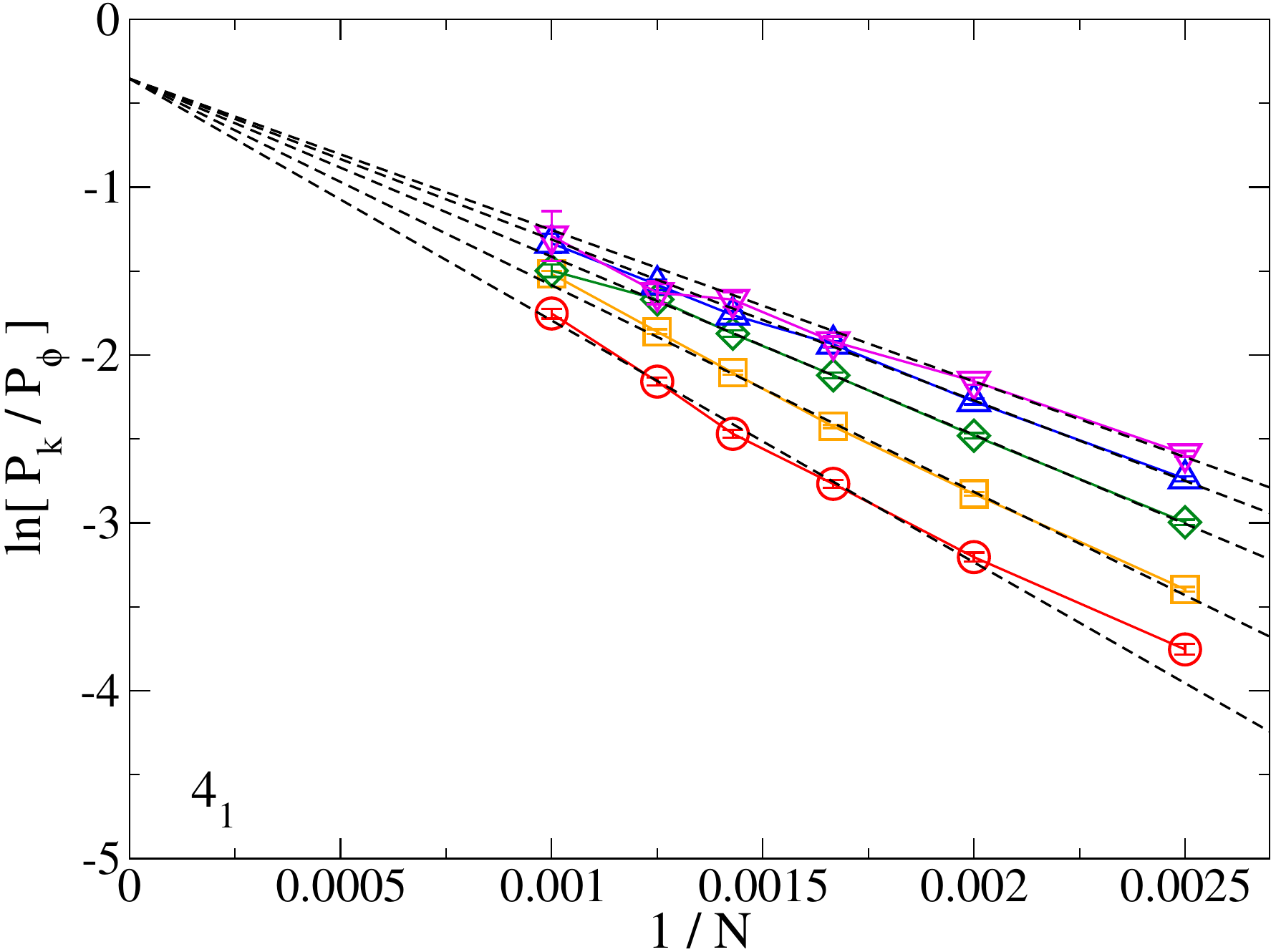}
\hskip 0.05\columnwidth
\includegraphics[width=0.65\columnwidth]{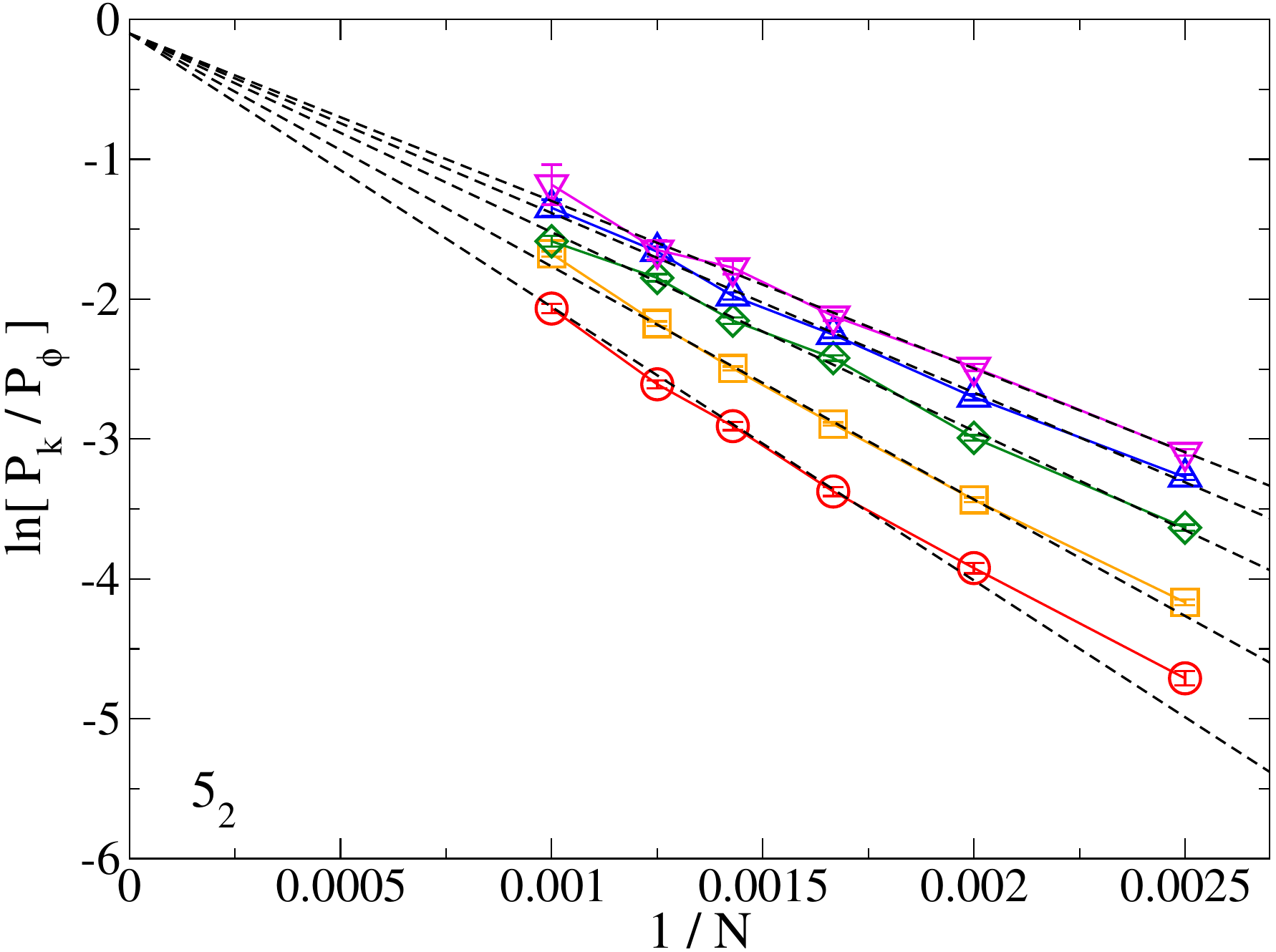}
\hskip 0.05\columnwidth
\includegraphics[width=0.65\columnwidth]{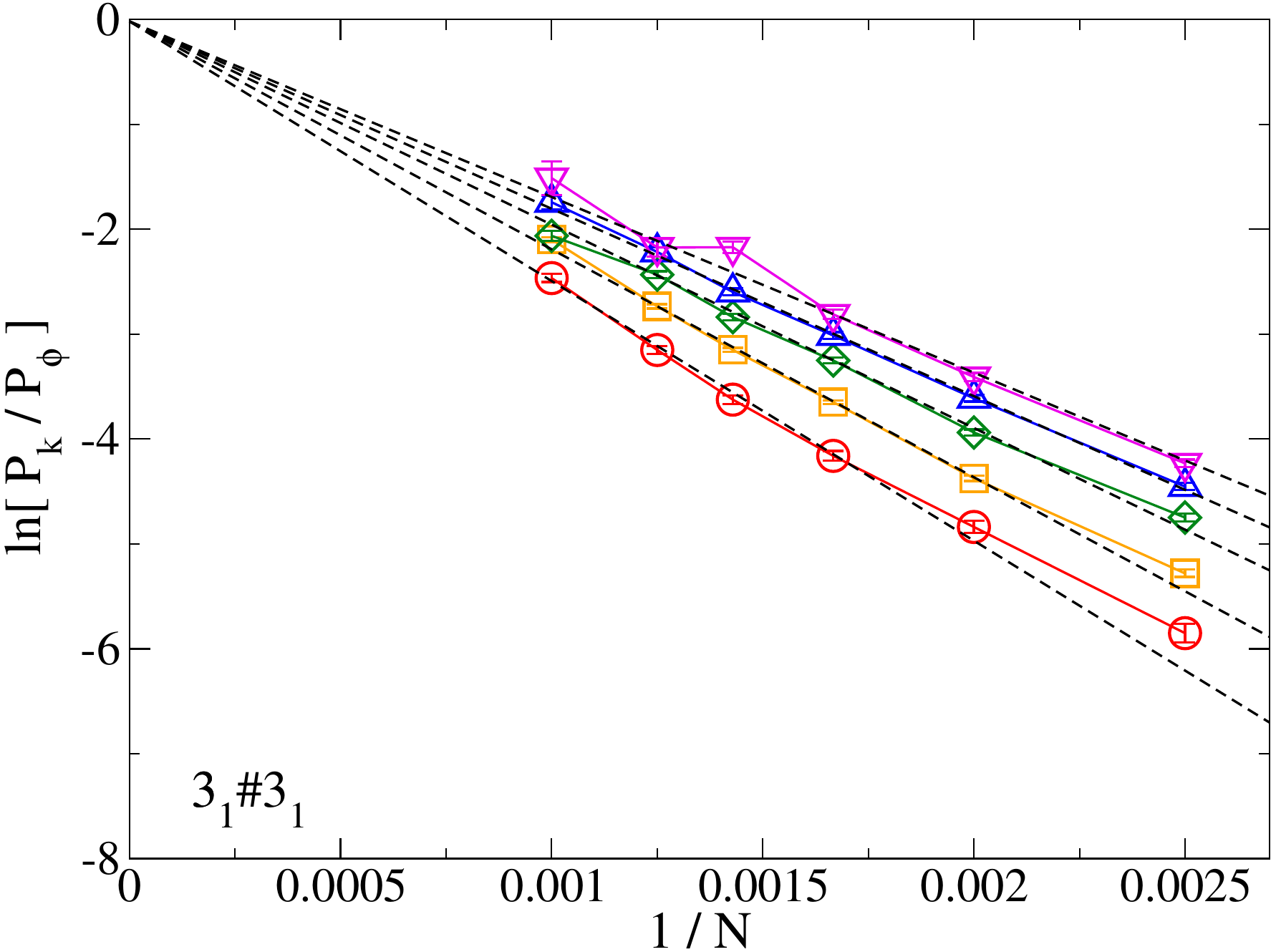}
\caption{Probability of occurrence of  a given knot type
($6$ cases shown, one for each panel) relative to the unknotting probability.
In each panel $5$ different temperatures are considered:
$1/T=0.37$ ({\large $\circ$}), $0.4$ ($\Box$), $0.44$ ({\large $\diamond$}), $0.48$ ($\vartriangle$), and
$0.52$ ($\triangledown$). }
\label{fig:5T}
\end{figure*}

At this point it is important to check how the scenario provided by
Fig.~\ref{fig:1N} changes upon varying the temperature of the globuli. At the
same time one should determine as accurately as possible the dependence
of the found topological free energy correction on both $n_c$ and $T$.
In the first panel of Fig.~\ref{fig:5T} we report,
for five different temperatures below $T_{\Theta}$,
the behavior with $1/N$ of the topological free energy correction
applying to the knot $3_1$.

A most remarkable feature in Fig.~\ref{fig:5T} is the fact that all the
plots at different temperatures appear to extrapolate linearly to the same asymptotic
relative frequency $\ln (A_{3_1}/A_\emptyset)$ for $1/N \to 0$.
Hence, the figure suggests that a fit of the form
\begin{equation}
\label{eq:fit}
\ln \left (\frac{A_{3_1}}{A_\emptyset} \right )- M_{3_1}(T) \frac{1}{N}
\end{equation}
can be applied to all data, with the same asymptotic ratio
and temperature-dependent slopes $M_{3_1}(T)$.
The dashed lines are indeed a global fit of the data based on this assumption,
confirming a nice consistency with it. The fact that the slope
of the linear fits, $M_{3_1}$, decreases in modulus with temperature
shows that the amplitude of the correction sensibly depends on
temperature, contrary to previous expectations~\cite{Baiesi:2011:PRL}.

A finite, temperature-independent asymptotic relative frequency
of a knot with respect to the unknot $\emptyset$ remains a
valid assumption also for the knots $4_1$, $5_1$, $5_2$, $6_1$, $3_1 \# 3_1$
(see the panels of Fig.~\ref{fig:5T}) and for all other
simple knots we have considered (not shown).
In the case of the two different knots with $n_c =5$ we see also that, while
the asymptotic relative frequencies are different, the slopes of the linear plots at
the same temperature are approximately the same. Thus, as
already verified in the case of Fig.~\ref{fig:1N}, the amplitude
of the topological correction depends on temperature and, at least to a good
approximation, on the minimal crossing number $n_c$ of the knot alone.

The asymptotic relative frequencies $\ln (A_k/A_\emptyset)$ extrapolated
from the above plots are reported in Table~\ref{tab:2}. These frequencies
are referring to knots which are delocalized into globules of
increasing size. It is thus legitimate to expect that, besides
being independent of temperature, they could be also universal
with respect to the type of lattice in which the knotted
configurations are realized. Remarkable universalities of
quantities related to asymptotic amplitudes have been detected
also in the study of the spectrum of knotted rings at $T\to\infty$
phase~\cite{Rensburg&Rechnitzer:2011:JPA,Baiesi:2012:PRE}.
A challenge posed by Table~\ref{tab:2} and its possible extensions
concerns the verification of such universality and the
possible link of the frequencies with topological invariants.
For sure our data seem to exclude that, if the amplitude ratios
depend on topological invariants, these could just reduce to
$n_c$ alone. The spectrum of ring polymers in the globular phase
is a remarkable case where topological polymer statistics puts all
knots an a sort of equal footing, independent of their prime or
composite character.

\begin{figure}[tb]
\includegraphics[width=0.95\columnwidth]{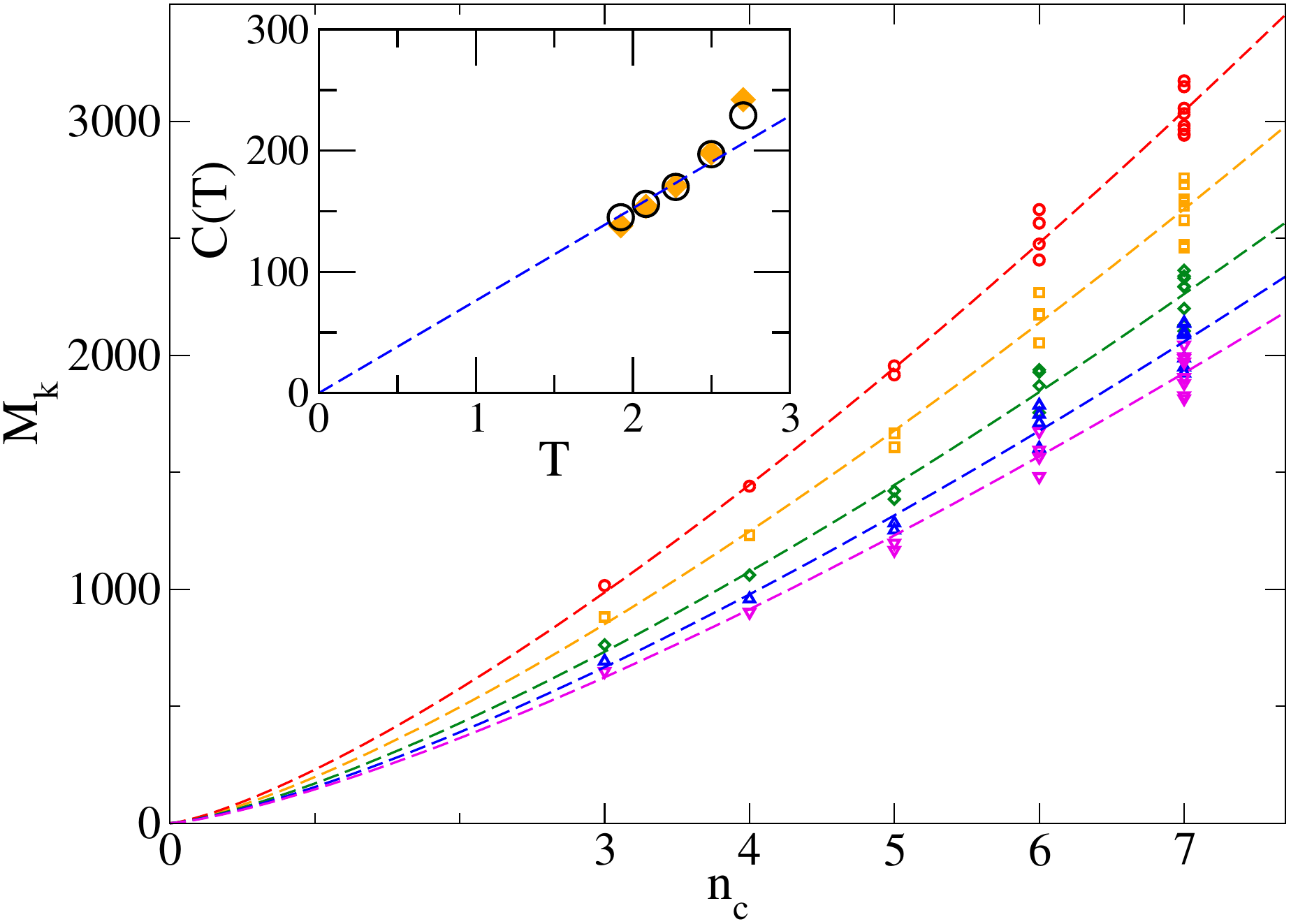}
\caption{Estimates of $M_k$ for knots up to $n_c=7$, at five temperatures
(see color/symbol code of previous figures).
Note that data with the  same $T$ and  $n_c$ are quite well clustered.
The dashed curves refer to fits of the form $\sim C(T) n_c^{1.33}$
(those in which  $\sim C(T) n_c^{a}$, with $a$ being a parameter free to vary
are similar).
Inset: Estimates of  $C$ as a function of temperature.
Filled diamonds are values determined  by not fixing the exponent $a$ in the fits,
while empty circles refer to estimates obtained by fixing
$a=1.33$ (i.e.~the weighted average of the $a$ estimates obtained  at the five
different temperatures considered).
The dashed line is a guide to the eye showing that $C(T)\sim T$ is indeed
an acceptable ansatz.
Fits of the most asymptotic black dots (not shown) give also
asymptotic $T\to 0$ extrapolations to a value $C(0)$ consistent with zero. 
}
\label{fig:Mk}
\end{figure}

What one can efficiently extract from the data at finite
$N$ reported in Figs.~\ref{fig:1N}-\ref{fig:5T}, is a tentative fitting form
for the topological correction as a function of temperature
and of $n_c$, assuming that these are the only parameters
determining it. If in the various plots one assumes an
asymptotic form as in Eq.~(3), an expression
of the form $M_k(T)= C(T) n_c^a$, factorizing the dependences,
 seems reasonable for the correction amplitude, as already assumed
in~\cite{Baiesi:2011:PRL}. This fitting form is applied globally in
 Fig.~\ref{fig:Mk} to the data collected for different knots and different temperatures.
The estimates, listed in Table~\ref{tab:1}, show that
an optimal determination for the exponent is $a \simeq 1.33$~\footnote{Previously 
$a \simeq 1.45$ was found~\cite{Baiesi:2011:PRL}. Moreover,
we have discovered that the value of $T=2.5$ quoted in Ref.~\cite{Baiesi:2011:PRL}
is not correct: simulations were actually run at $T=2$, which explains
some discrepancies found in estimates of $C(T)$.}.
We have thus fitted again data with the one-parameter form $C(T) n_c^{1.33}$ and obtained
similar values of $C(T)$, see Table~\ref{tab:1}.
In the inset of Fig.~\ref{fig:Mk} we report the determinations of the amplitude $C(T)$
resulting from the last fits. 
Even if the temperature range covered by our determinations remains limited,
there is evidence that the postulated $C(T)$ could vary
linearly with temperature and could approach zero for $T\to 0$.
An important fact are the relatively large values
of $C$ determined for all temperatures considered. Such large values
show that the topological correction is quite substantial
for not too large globules. Notice that both $C(T)$ and the exponent $a$ are just intended
to provide a convenient fitting for the quantities $M_k(T)$  reported in  Table~\ref{tab:2}.
In particular, one should not assign a particular meaning to the value of the exponent $a$.

\begin{table*}[!t]
\begin{center}
{ \footnotesize
\begin{tabular}{l  l  l l l l l l l l }
    \hline
knot\ \ \  & $M'(0.48)$\ \ \  \ \ & $\ln(A_k/A_{\emptyset})$  & $M(0.37)$ & $M(0.4)$ & $M(0.44)$ & $M(0.48)$ & $M(0.52)$  & $\overline{l}$ & 
Eq.~(10)\\
\hline
$3_1$ & 714(21) & 0.39(0.12) & 1020(73) & 885(71) & 765(59) & 696(57) & 652(54) & 49(1) & 66(11) \\
\hline
$4_1$ & 929(15)& -0.36(0.14) & 1434(89) & 1224(85) & 1055(72) & 953(68) & 897(64) & 67(2) & 80(12)\\
\hline
$5_1$ & 1260(30)& -0.74(0.18) & 1910(120) & 1600(120) & 1380(100) & 1250(90) & 1170(90) & 77(2) & 94(14)\\
$5_2$ & 1290(40)&-0.10(0.17) & 1960(120) & 1670(110) & 1420(90) & 1280(90) & 1200(80) & 85(6) & 96(14)\\
\hline
$6_1$ & 1660(80)& -0.73(0.27) & 2400(190) & 2050(180) & 1750(150) & 1600(140) & 1480(130) & & \\
$6_2$ & 1690(40)& -0.41(0.22) & 2570(160) & 2180(150) & 1880(130) & 1720(120) & 1570(110) & 105(6) & 114(16)\\
$6_3$ & 1795(85)& -0.78(0.42) & 2700(300) & 2310(280) & 1970(240) & 1780(230) & 1630(210) & & \\
$3_1$\#$3_1$  & 1805(35)& 0.00(0.26) & 2490(180) & 2190(170) & 1950(150) & 1800(140) & 1680(130) & & \\
\hline
$7_1$&& -2.17(0.53) & 2980(390) & 2460(360) & 2110(320) & 1930(290) & 1830(270) & & \\
$7_2$&& -1.24(0.49) & 2980(360) & 2590(330) & 2210(290) & 2000(270) & 1890(250) & & \\
$7_3$&& -1.56(0.43) & 2920(320) & 2450(290) & 2110(250) & 1940(240) & 1800(210) & & \\
$7_4$ & 2260(60)& -1.94(0.50) & 3040(350) & 2650(340) & 2310(300) & 2080(290) & 1890(250) & & \\
$7_5$&& -0.95(0.35) & 3050(250) & 2670(240) & 2300(210) & 2100(190) & 1980(180) & & \\
$7_6$&& -0.75(0.28) & 3160(200) & 2740(190) & 2350(170) & 2110(160) & 2000(140) & & \\
$7_7$&& -1.50(0.43) & 3150(310) & 2740(300) & 2350(260) & 2120(250) & 2030(220) & & \\
$3_1$\#$4_1$ & 2110(55) & -0.60(0.44) & 2960(320) & 2650(300) & 2310(260) & 2150(250) & 2000(220) & & \\
    \hline
\end{tabular}
}
\end{center}
\caption{Estimates of the knot-dependent constants for some simple knots
in globules. The column  $M'$ refers to parameters from fits of $\ln( P_k/P_\emptyset )$ 
vs. $1/N$ only at $1/T=0.48$,
while the following columns are from fits assuming that all data at different $T$
converge to the same asymptotic value $\ln( A_k/A_\emptyset )$,  as in Fig.~\ref{fig:5T}.
The last two columns contain the 
estimates of the typical globule length  $\overline{l}$  for some knots for which we had clear peaks in the
distributions $P(l_1)$ for the competition of Fig.~\ref{fig:sl}(b). The column $\overline{l}$ shows direct
estimates (positions of the peaks in the limit of long chain lengths, not shown), 
to be compared with predictions from Eq.~(10) in the last column
(we used $\sigma=-0.96(5)$, see Ref.~\cite{Baiesi:2014:PRE}, and values in the column $M(0.48)$).
}
\label{tab:2}
\end{table*}

\section{The role of $n_c$ in the thermodynamics of the globule}

Since globules in experiments are necessarily finite, the above
findings about the topological correction to the globule free energy
suggest that the invariant $n_c$ could play an important
role in the thermodynamic behavior of collapsed knotted rings.
In order to shed light on this role, a natural way is to apply
to the globule geometrical constraints which can interfere with
topology, and in particular with the topological invariant under
examination. The response to such constraints could reveal
interesting ways in which topology controls the behavior of
the globule. A further possibility is that of considering geometrically
constrained situations which allow a direct check of the presence and
effects of the postulated free energy correction. In this section we
pursue the first ways of analysis.

Since now we are interested in configurations
with fixed topology, we use the BFACF method described in the previous section,
and we tune the parameters to sample chains longer than those obtained by PERM
for unrestricted topology. In this way, we arrive at lengths $N$ up to $7000$.

\begin{figure}[bt!]
\begin{tabular}{ccc}
\includegraphics[width=0.4\columnwidth]{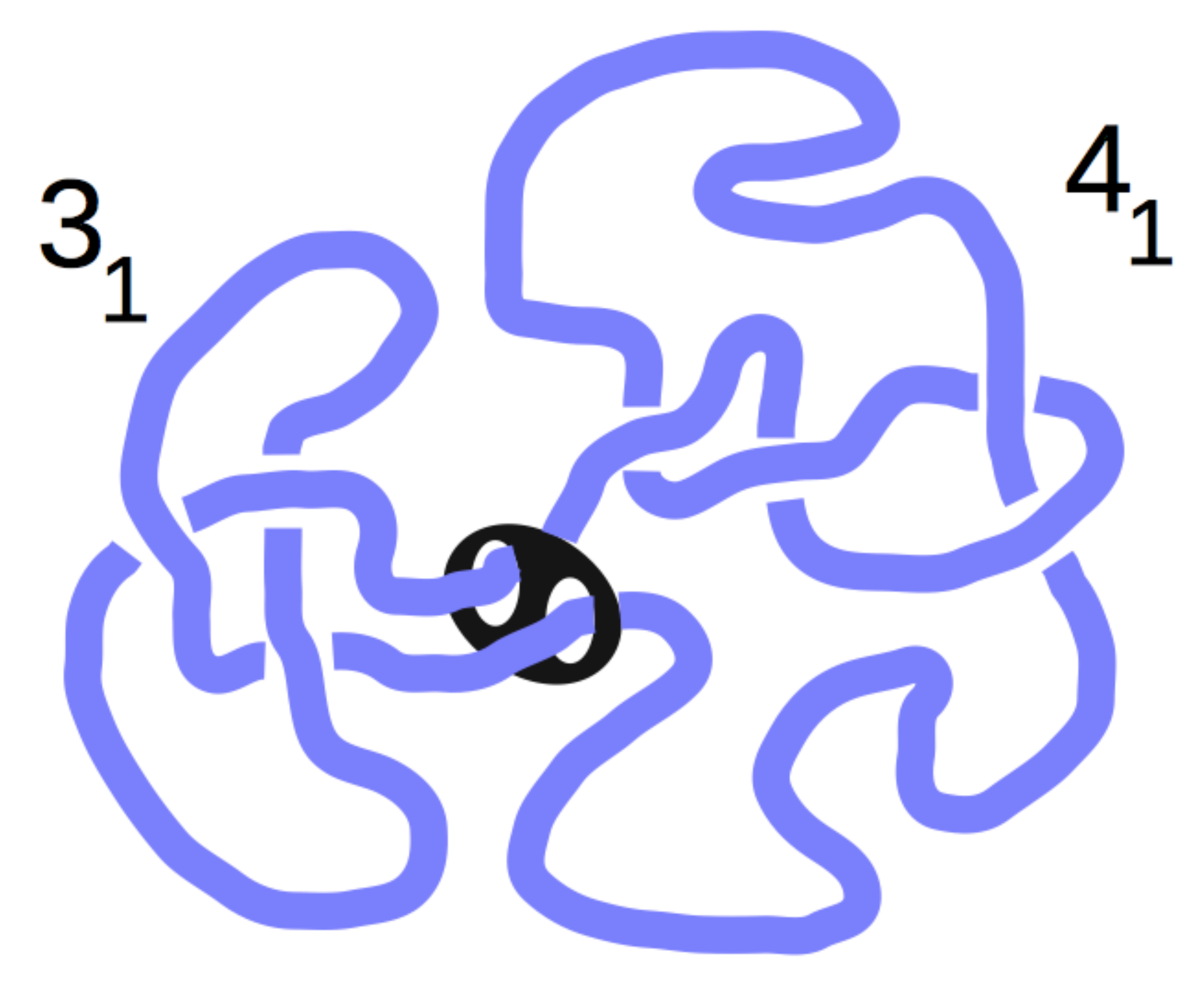} &  &
\includegraphics[width=0.47\columnwidth]{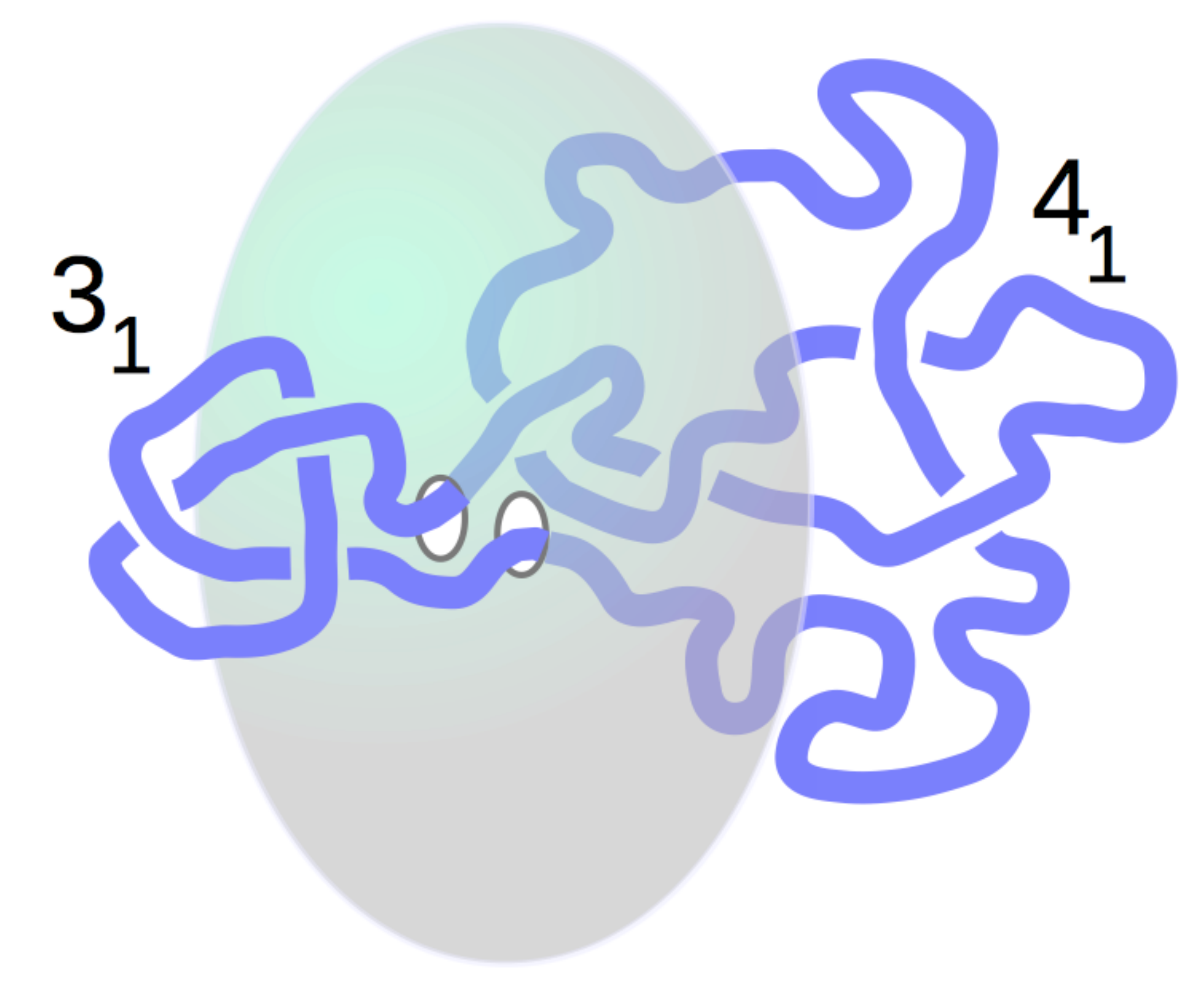} \\
(a)  & \phantom{( )}&  (b)  
\end{tabular}
\caption{Sketch of a polymer ring with a slip-link separating two loops, the first holding
a $3_1$ knot and the second a $4_1$. Each knot is confined in its loop because the holes in the slip-link
are narrow enough to prevent the passage of a blob containing more than one monomer.
However, in this way the loop lengths may vary.
In case (a) the two loops interact as a single globule, while in (b) they form two
separate globules because the slip-link is in an impenetrable wall.}
\label{fig:sl}
\end{figure}

\begin{figure}[bt!]
\includegraphics[width=0.9\columnwidth]{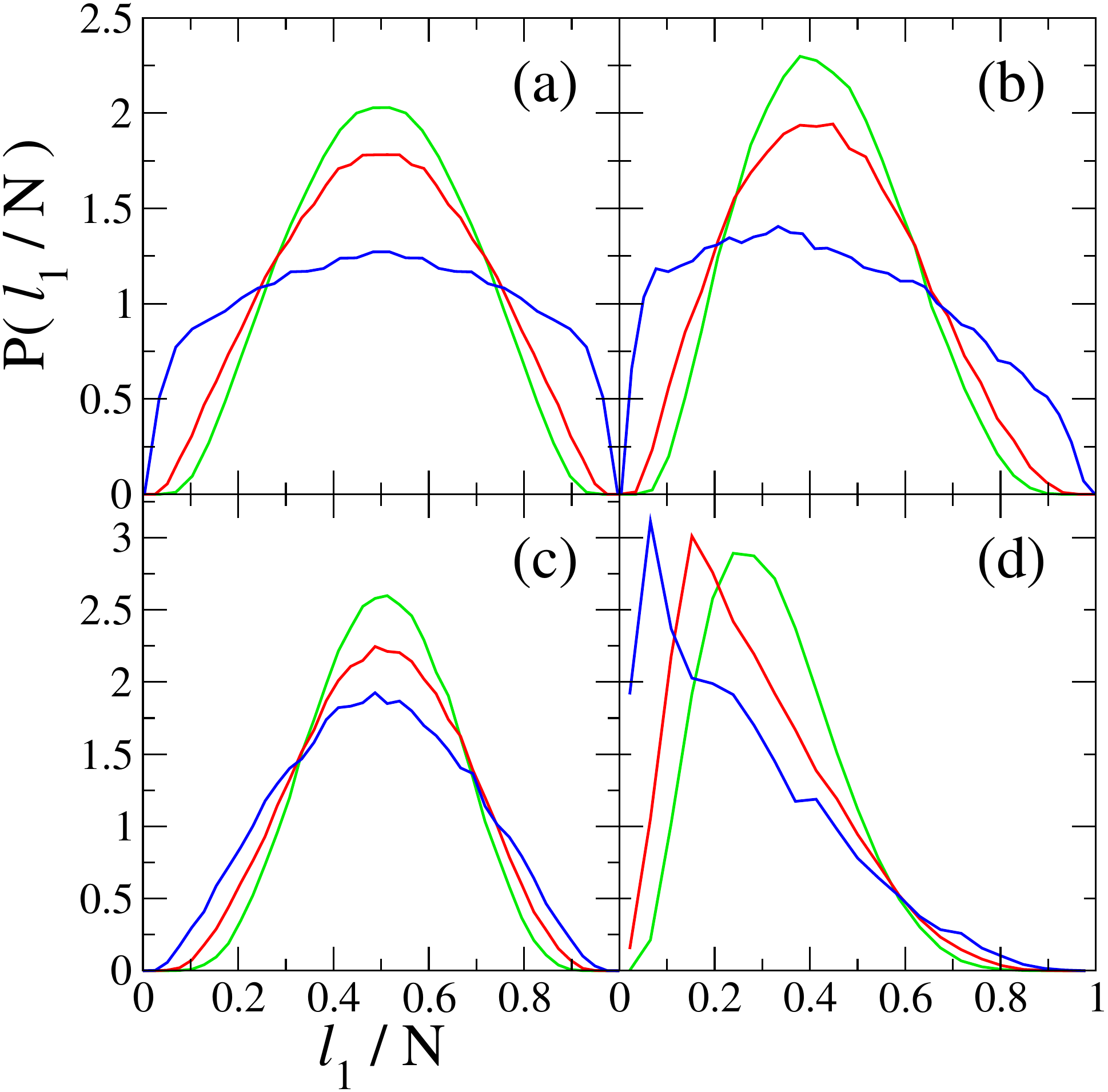}
\caption{Probability distributions of the fraction of
length of one of the two loops in a slip-link competition of knots [Fig.~\ref{fig:sl}(a)], for
different ring length:
(a) $3_1$ vs.~$3_1$, for $N=500$ (green), $1000$ (red), $4500$ (blue);
(b) $3_1$ vs.~$4_1$, for $N=500$, $1000$, $5000$;
(c) $6_1$ vs.~$3_1\#3_1$, for $N=1000$, $2000$, $4000$;
(d) $3_1$ vs.~$7_1$, for $N=500$, $2000$, $7000$.
}
\label{fig:Pl}
\end{figure}

We first consider the effect of a slipping link on the statistics
of the globule. The slipping link, sketched in Fig.~\ref{fig:sl}(a) is such to divide
the ring in two loops, of length $l_1$ and $l_2=N-l_1$. Each loop
has a given knot, respectively $k_1$ and $k_2$, and the slipping
link is narrow enough to prevent the translocation of the knots
from one loop to the other. This is the interference between geometry
and topology mentioned above. It should be stressed that the
two loops are attractively interacting in this set up, so that
they are forming a single globule.

It is interesting to analyze
the fluctuations in the lengths of the two loops in equilibrium,
and to verify in which way they possibly depend on the two knots
in the loops. To this purpose we use the quasi-canonical simulation
method preserving topology described in Section {\em Relative frequencies and free energies of knotted globuli}.
In Fig.~\ref{fig:Pl} we report the probability density function $P(l_1/N)$
of the fraction of total backbone taken on average by loop 1,
for different $N$ and for different competing knots $k_1$ and $k_2$.
All runs were rather extensive and performed at $T=2$, below the $\Theta$ point. 
In Fig.~\ref{fig:Pl}(a) we see the competition between two $3_1$ knots.
The tendency of the distribution to become
broader and flatter with increasing $N$ indicates that
$l_1$ undergoes broad fluctuations, which grow proportional to $N$ itself.
The same tendency is observed in other knot competitions displayed in
Fig.~\ref{fig:Pl}.  A case of $3_1$ in loop 1 competing with
$4_1$ in loop 2 is reported in Fig.~\ref{fig:Pl}. In this case
there is an increasing skewness of the distribution for increasing $N$,
showing that configurations in which the loop 1 is shorter than loop 2
are favored. This dominance occurs again in the presence of
broad fluctuations, as indicated by the fact that the distribution broaden
with increasing $N$.
The distributions become more skewed in general if
$n_{c1}$ is much different from $n_{c2}$, as one can see
for a $3_1$ vs. $7_1$ competition in Fig.~\ref{fig:Pl} (d)], while they
remain (approximately) symmetrical even for different knots
if $n_{c1}= n_{c2}$ [case of a $6_1$ vs.~$3_1\#3_1$ shown in Fig.~\ref{fig:Pl}(c)].

\begin{figure*}[tb]
\includegraphics[angle=0,width=0.9\textwidth]{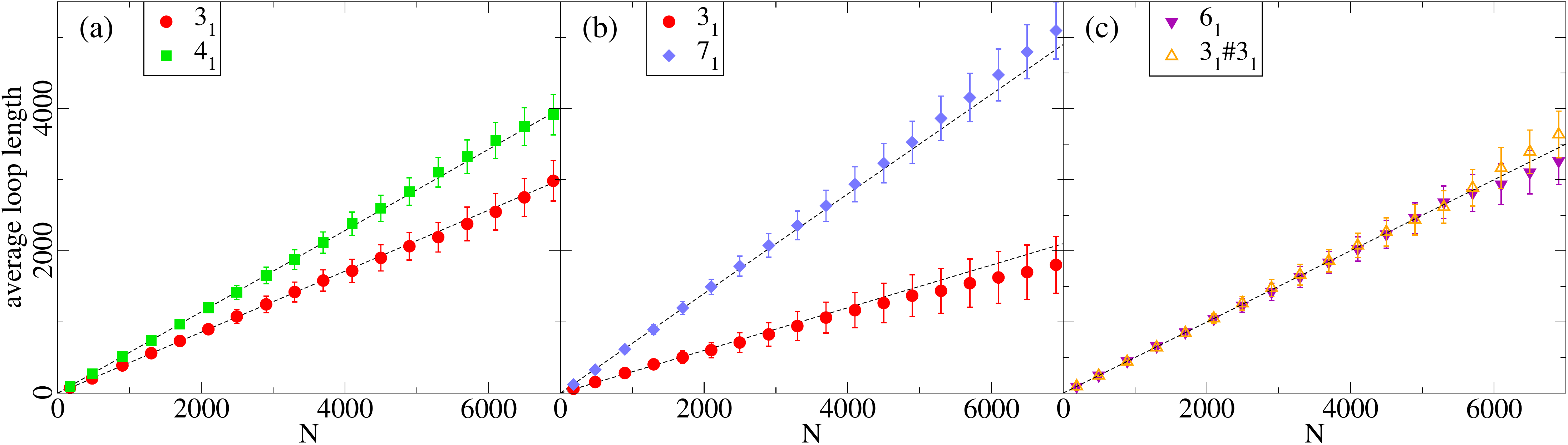}
\caption{Average length of each of the two loops in which the globule is partitioned by the slip-link, see
Fig.~\ref{fig:sl}(a).	The knots hosted by the two loops are
(a) $3_1$ and  $4_1$ ;
(b) $3_1$ and  $7_1$;
(c) $6_1$ and  $3_1\#3_1$.
Straight lines have slopes predicted by Eq.~(4). }
\label{fig:ll}
\end{figure*}

Given the reported increasingly broad distributions of loop lengths, it is
not straightforward to predict the scaling of their averages $\mean{l_1}$  
and $\mean{l_2}$ with $N$. However, the plots of $\mean{l_1}$ and $\mean{l_2}$
 as a function of $N$ reported in Fig.~\ref{fig:ll}
reveal a surprisingly simple linear behavior.
For example, for the case $k_1=3_1$ vs $k_2=4_1$
in Fig.~\ref{fig:ll}(a) we observe that $\mean{l_1}$ and $\mean{l_2}$
approximately grow proportional to $\frac 3 7 N$ and $\frac 4 7 N$,
respectively (see the straight lines).
This suggests that $n_c$ is the controlling parameter in the competition.
Another example of competition is a $3_1$ against $7_1$,
as reported in Fig.~\ref{fig:ll}(b). Here again we have seen that the fluctuations
of $P(l_1/N)$ are broad, yet the average loop lengths are determined on the basis of
the minimal crossing numbers of the competing knots,
 as $\frac 3 {10} N$ and $\frac 7 {10} N$.
A further confirmation of the fact that $n_c$ alone determines the
loop statistics, is given by the competition $3_1 + 3_1$ vs $6_1$.
In this case [Fig.~\ref{fig:ll}(c)] we see that the symmetry of $P(l_1/N)$ shown
in  Fig.~\ref{fig:Pl}(c)
is also  well reflected by the behavior of the average loop lengths
$\mean{l_1}\simeq\mean{l_2}\simeq N/2$.

On the basis of the above simulations one can postulate that
the average loop lengths are determined according to
\begin{equation}
\label{l1l2}
\mean{l_i}=  \frac{n_{ci}}{n_{c1} +n_{c2}} N
\end{equation}
where $n_{ci}$ is the minimal number of crossings in loop $i$.
This is a deceptively simple law which could, however, be hardly guessed
a priori. Of course, the provided evidence that this law applies is
limited by simulation capabilities. Like in the case of the results
presented in the previous section, there is the risk that the exclusive
and sharp role postulated for $n_c$ is valid only to a very good
approximation. For example, looking at the data for the $3_1\#3_1$
vs $6_1$ competition, the circumstance that the two loops are going
to a fully symmetrical equilibrium for large $N$ can be reasonably
guessed, rather than given for certain. On the other hand, even in
the perspective of establishing approximate laws, the results presented
here appear remarkable. Thinking to the enormous geometrical
intricacy in which the essential topological crossings are hidden
within the globule configurations, one would guess at first sight
that the role of $n_c$ should not be so important.

\section{Effects of the topological correction in translocation}

A way to make direct contact with the ansatz for the knotted
globule free energy presented in the previous section is to replace
the slipping link with an impenetrable wall presenting a hole through which
the knotted ring is passing (Fig.~\ref{fig:sl}). In this setting the two loops would
be not interacting with each other, and would constitute independent
knotted globuli just competing for backbone length, each one with its own surface 
exposed to the solvent.  This resembles the translocation of a globular ring polymer through
a membrane or a solid state nanopore (though in our setup the hole does not allow a
full translocation, preserving the knots in each loop). Situations like this should be
in principle within the reach of experimental investigation nowadays,
thanks to progresses in nanophysics and nanotechnology.

Important here is the fact that the free energy
of the whole system reduces to the sum of the free energies of the two
globuli, due to their independence. Moreover, in the limit of one short loop, the complementary
long loop acts just as a monomer reservoir. Below we show that $M_k$
is related to the typical chain length selected by the
globule in that limit and to the length of ideal knots.

\begin{figure*}[t!]
\includegraphics[width=0.9\textwidth]{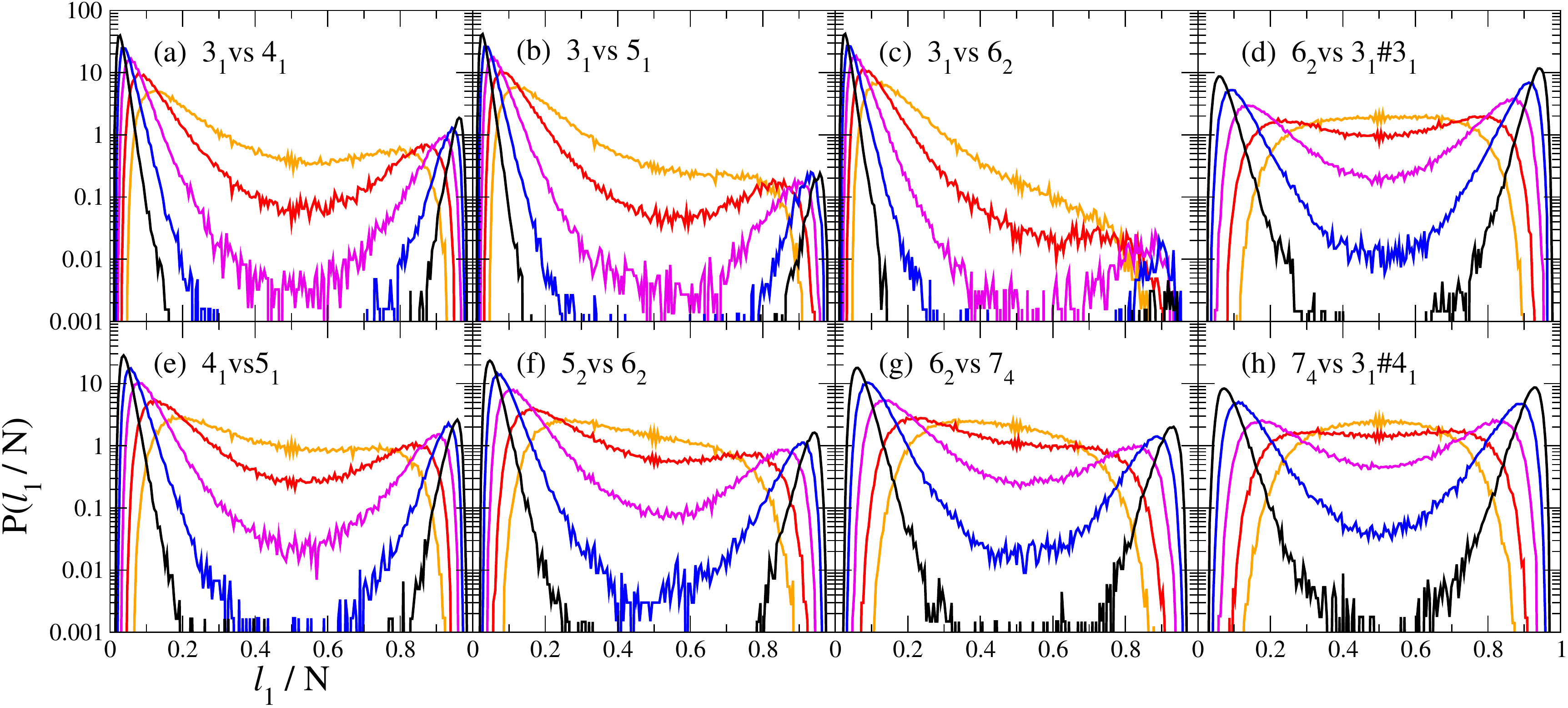}
\caption{Probability distributions of the relative
length of one of the two loops in a competition of two globular knots separated by a wall [Fig~\ref{fig:sl}(b)], 
for different knot pairs.
In each case five different $N$ values ($500$, $700$, $1000$, $1400$, and $2000$)  are considered.
The sequence (a)-(b)-(c) refers to the	knot $3_1$ vs a progressively more complex knot: 
one can note an increasing unbalance in the distributions.
The sequence (a)-(e)-(f)-(g) is for progressively more complex knots, 
showing that the minimum between the two maxima of the distributions occurs for longer $N$'s if the hosted
knots become more and more complex; this is due to a larger minimal knot size.
Finally, (d) and (h) are examples of knots competing with other knots that have
the same minimal number of crossings $n_c$ but are otherwise topologically
different: the symmetry of the distributions confirms that $n_c$ is
a good number to look at when knots are competing in the globular phase.
}
\label{fig:Pl8}
\end{figure*}

The logarithmic weight (or minus dimensionless free energy) of the first globule
of length $l_1$, defined as $\varphi_{k_1}(l_1)  \equiv -{F_{k_1}(l_1)} / {k_B T}$ 
(where $F_{k_1}$ is the total loop free energy), is written as
\begin{equation}
\label{F}
\varphi_{k_1}(l_1)  =  \ln A_k + \kappa l_1 + \sigma l_1^{2/3}
+(\alpha-2)\ln l_1 - \frac{M_{k_1}}{l_1}
\end{equation}
where $\kappa$, $\sigma$, and $M_{k_1}$ are functions of $T$ 
(we are omitting further knot-independent corrections to scaling).
Thus, for the translocating ring one can write the probability of the first loop length
when competing with a knot $k_2$,
\begin{eqnarray}
P_{k_1,k_2}(l_1) &\propto& \exp\left[\varphi_{k_1}(l_1) + \varphi_{k_2}(N-l_1)\right]
\end{eqnarray}
With the free energy form given in Eq.~(5), in the case of
$n_{c1}=n_{c2}$ one would expect $M_{k_1}\approx M_{k_2}$, i.e. a symmetric $P$, with two
equivalent maxima determined by the surface tension term.
This is what we observe by analyzing our simulations,
for example in the case $6_1$ vs $3_1\#3_1$, as shown in Fig.~\ref{fig:Pl8}(d),
or $7_4$ vs $3_1\#4_1$ (Fig.~\ref{fig:Pl8}(h)),
where we plot distributions of the rescaled variable $x = l_1 / N$, 
for a better comparison of data with different $N$.
The plots show that for increasing $N$ there is a
competition between two stable minima of the free energy,
corresponding to states in which one of the globules
takes the largest part of the backbone length.

When the same type of simulation is performed for a case
like $3_1$ vs $4_1$ (Fig.~\ref{fig:Pl8}(a)), the scenario changes:
$P$ becomes very asymmetric, showing a pronounced, dominant
peak for configurations in which the globule with $k=4_1$
takes most of the backbone length. The peak for opposite
configurations with dominating $3_1$ is considerably depressed
and almost disappears in comparison. Panels (b)-(c) and (e)-(g)
of Fig.~\ref{fig:Pl8} show similar cases. The message is very clear:
the dominance of only one peak is a consequence of topology
and thus should be ascribed to the topological correction.

\begin{figure}[tb]
\includegraphics[angle=0,width=0.9\columnwidth]{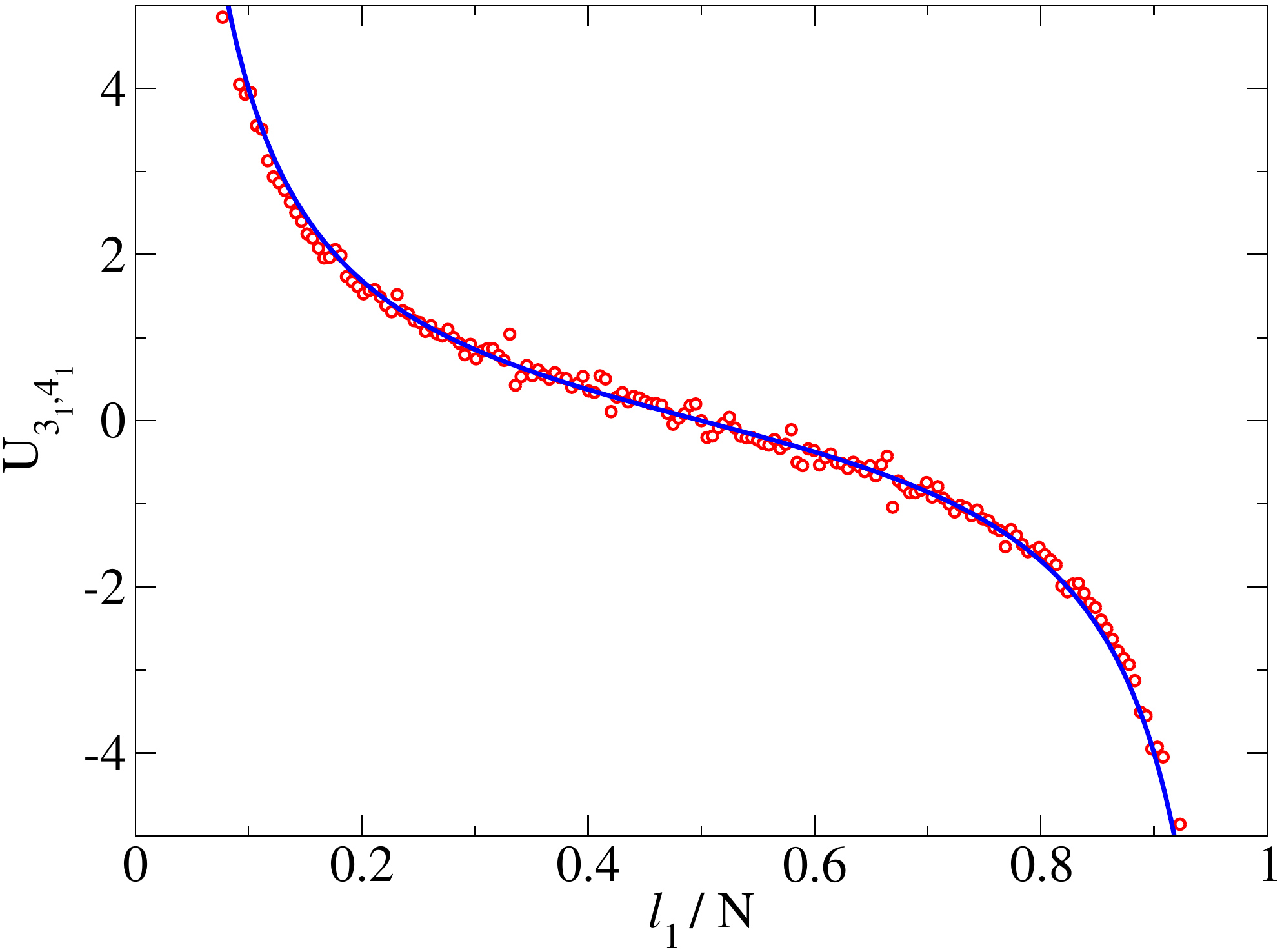}
\caption{Free energy differences vs. the rescaled loop length for a $3_1$ competing with  a $4_1$
in the configuration of Fig.~\ref{fig:sl}(b) for $N=500$.
Points are data in the form $\ln [  P_{k_1,k_2}(l_1)/P_{k_1,k_2}(N-l_1) ]$ while
the solid curve is a plot of $U_{k_1,k_2}(l_1;N)$ 
fitted with Eq.~(\ref{Delta}) through these data.
}
\label{fig:Delta}
\end{figure}

To check whether these knot-plus-wall simulations yield results compatible with the data from
isolated polymers, one may consider combinations of the data
that depend exclusively on the expected topological correction.
For instance, one can consider the weight ``unbalance'' (or free energy difference)
\begin{eqnarray}
\label{Delta}
U_{k_1,k_2}(l_1;N) &\equiv& \ln [  P_{k_1,k_2}(l_1)/P_{k_1,k_2}(N-l_1) ]\nonumber\\
&\simeq&  \varphi_{k_1}(l_1) + \varphi_{k_2}(N-l_1) \nonumber\\&&
-\left[ \varphi_{k_1}(N-l_1) + \varphi_{k_2}(l_1)\right]
\nonumber \\
&\simeq& (M_{k_1}-M_{k_2})\frac{N-2 l_1}{l_1(N-l_1)}\nonumber\\
&=& \frac{M_{k_1}-M_{k_2}}N \frac{1- 2 x }{x(1-x)}
\end{eqnarray}
One can note that only a difference $\Delta M = M_{k_2}-M_{k_1}$ determines this distribution.
In Fig.~\ref{fig:Delta} we show a plot of $\ln [  P_{k_1,k_2}(l_1)/P_{k_1,k_2}(N-l_1) ]$
and of its fit $U_{k_1,k_2}(l_1;N)$ (for $T=1/0.48$, $N=500$, $k_1=3_1$, $k_2=4_1$) vs
$x=l_1/N$. 
The value $\Delta M \simeq 224$  from the fit is reasonably consistent with the determinations 
$M_{4_1}-M_{3_1}\simeq 263$ and $M'_{4_1}-M'_{3_1}\simeq 215$ based on the
results for knot frequencies (which were summarized in Table~\ref{tab:2}).
This agreement is found for other knot pairs as well, see Table~\ref{tab:3} where, 
for each knot combination and for both $N=500$ and $N=1000$, we list two estimates:
$\Delta M_0$ from fits of all data and
$\Delta M_1$ from fits of data with $0.2\le x \le 0.8$.
The estimate $\Delta M_1$ is introduced because it could be less sensible to 
knot-independent corrections to scaling, which should be relevant  when one
of the loops becomes too short.
From Table~\ref{tab:3} one can see that both estimates are quite consistent with a null value for cases
$n_{c1}=n_{c2}$ or with values coming from the data for isolated knots.

Thus, the results concerning the competition of globules in
a translocation set up give further independent confirmation
of the presence of the topological free energy correction
and show at the same time its importance for the physics of
this process. The outlined intriguing scenario calls for possible
fundamental justifications or interpretations of the presence of
this correction.

\begin{table*}[!bt]
\begin{center}
\begin{tabular}{ c c  l l  c c  c c }
    \hline
 & & & & \multicolumn{2}{c}{$N=500$} & \multicolumn{2}{c}{$N=1000$}\\
$k_1$ & $k_2$ &$M_{2}-M_{1}$ &$M'_{2}-M'_{1}$ & $\Delta M_0$ & $\Delta M_1$  & $\Delta M_0$ & $\Delta M_1$ \\
\hline
$3_1$ & $4_1$ & 263 & 215(36) & 224(3) & 225(3) & 224(6) & 189(32)\\
$3_1$ & $5_1$ & 560 & 546(48) & 385(2) & 408(3) & 401(6) & 469(30)\\
$3_1$ & $6_2$ & 1018 & 979(56) & 735(5) & 763(3) & 1469(10) &	1526(11)\\
$4_1$ & $5_1$ & 297 & 331(42) & 163(2) & 183(3) & 191(3) &	253(16)\\
$5_2$ & $6_2$ & 428 & 401(71) & 297(3) & 298(2) & 322(4) & 392(10)\\
$6_2$ & $7_4$ & 383 & 570(95) & 286(4) & 292(3) & 308(3) &	362(5)\\
$6_2$ & $3_1$\#$3_1$ & 75 & 112(68) & -53(2) & -48(2) & -53(3) &	-9(5)\\
$7_4$ & $3_1$\#$4_1$ & 24 & -152(112) & -12(3) & -14(2) & -3(3)  & 14(4)\\
    \hline
\end{tabular}
\end{center}
\caption{Fits of $\ln P(l/N)/P(1-l/N)$, for several knots at $1/T = 0.48$:
here $\Delta M_0$ is from the fit of the whole data while
$\Delta M_1$ is from only data with $0.2\le l/N \le 0.8$.
Each value is computed for the global ring lengths $N=500$ and $N=1000$.
The values $M_{2}-M_{1}$ and $M'_{2}-M'_{1}$ are determined with
the simulations described in the previous section, see Table~\ref{tab:2}}
\label{tab:3}
\end{table*}

Referring to our ansatz for the globule free energy, Eq.~(5),
the condition $\frac{\partial}{\partial l_1}[\varphi_{k_1}(l_1)+\varphi_{k_2}(N-l_1)] = 0$ 
determining the most probable value of $l_1$
becomes independent on $\varphi_{k_2}$ if one looks for small values of $l_1$ with long $N$:
\begin{equation}
\frac{2}{3} \sigma l_1^{-1/3}+(\alpha-2) l_1^{-1} + M_{k_1}  l_1^{-2} \simeq 0
\end{equation} 
The solution $\overline{l_1}$ is indeed a relatively short length. This
in spite of the fact that we assume still validity of the
form in Eq.~(5) for the first globule. The last equation gives for
$M_k$ a leading behavior
\begin{equation}
M_{k_1}(T) \simeq   -\frac{2}{3} \sigma(T) \overline{l_1}^{5/3} -(\alpha-2)\overline{l_1}
\end{equation}
(we recall that $\sigma<0$).
The second term on the righthand side is a small correction, and the relation
may be approximately inverted to express the typical length $\overline{l_1}$ as a function of $M_{k_1}$,
\begin{equation}
\label{l_1_bar}
 \overline{l_1}(T) \simeq \left[-\frac {3 M_{k_1}(T)}{2 \sigma(T) }\right]^{3/5}
\end{equation}
Recent results~\cite{Baiesi:2014:PRE} give estimates of  $\sigma(T)$ in the range $0.32 \le 1/T \le 0.5$.
From these results, at the inverse temperature $1/T=0.48$ we interpolate $\sigma= - 0.96(5)$, which can be used
to check that the predictions of (10) are fairly consistent with 
direct estimates of $\overline{l_1}$ from the maxima of $P(l_1)$ , see the last two columns of Table~\ref{tab:2}.
There could be a bias of the indirect estimates toward larger values, which might come from
knot-independent corrections to scaling. However, the general trend follows closely that of the 
 $\overline{l_1}$ of probability maxima.

\begin{figure*}[tb]
\includegraphics[angle=0,width=0.9\columnwidth]{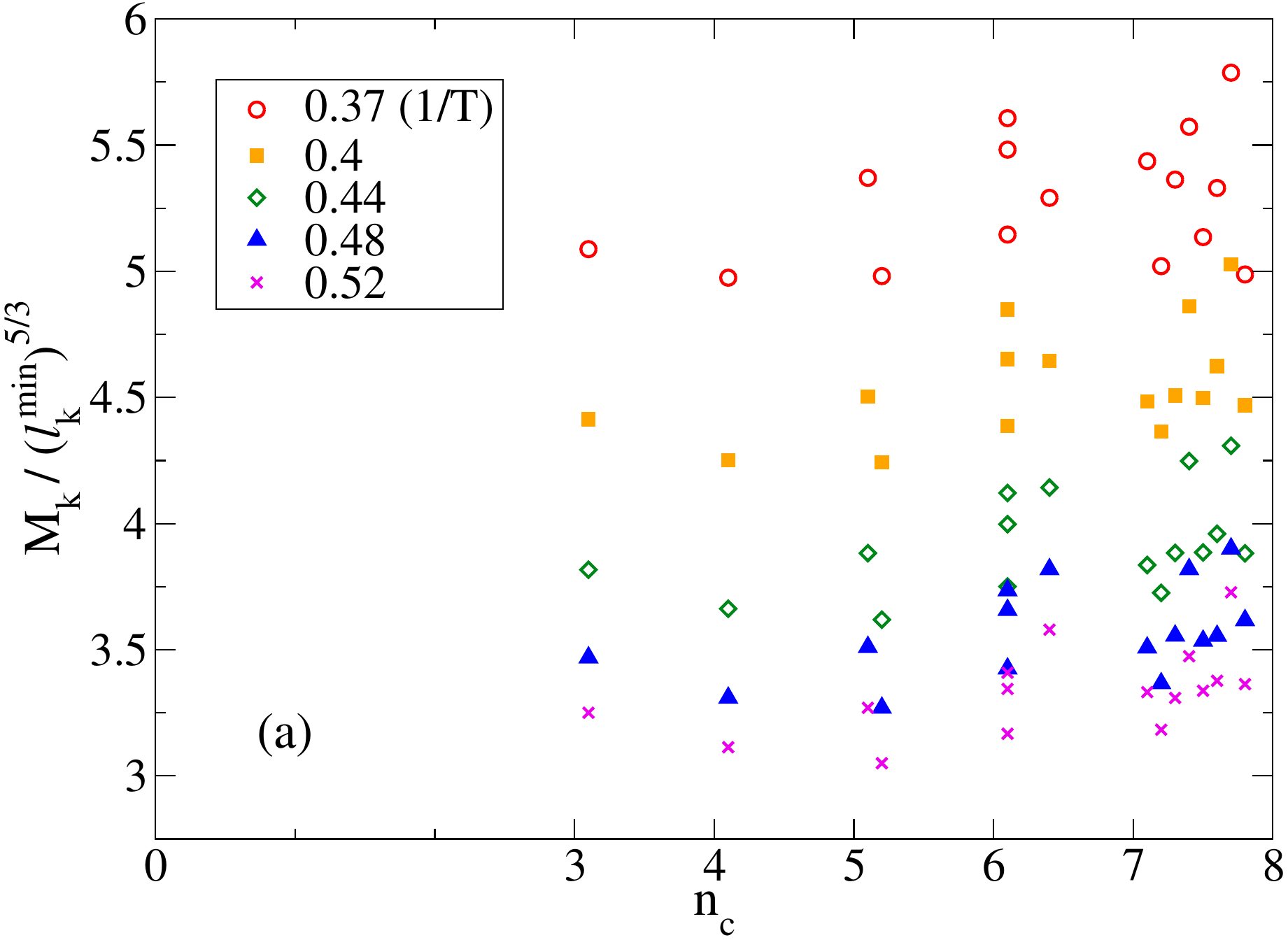}
\hskip 3mm
\includegraphics[angle=0,width=0.9\columnwidth]{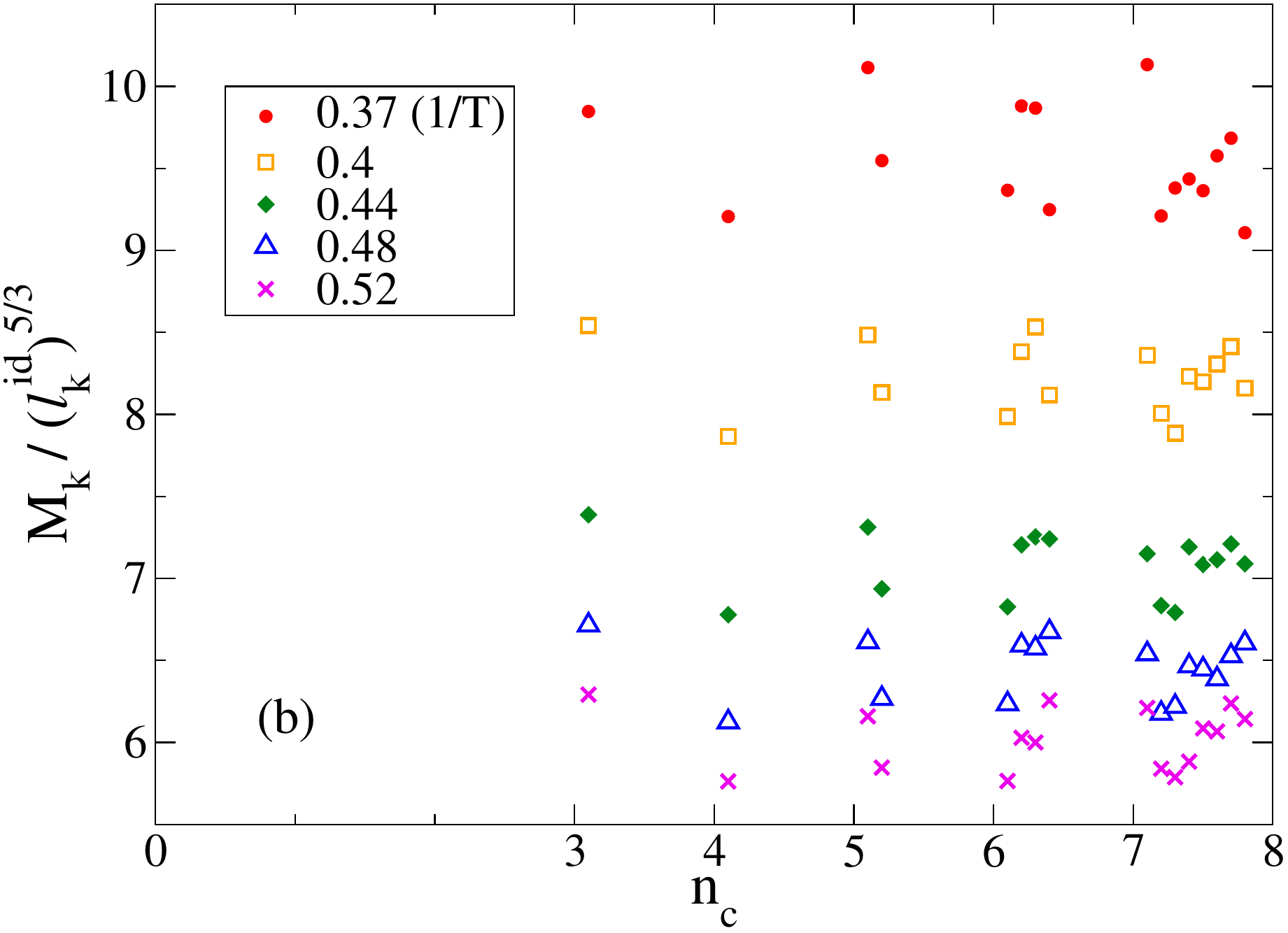}
\caption{
Ratio between correction-to-scaling constants and ideal knot lengths 
to the power $5/3$ (top panel for minimal lengths on the cubic lattice, bottom panel for
ideal lengths of knotted curves in three dimensions), for different temperatures.
Abscissas are equal to the knot crossing number plus $1/10$ of the knot index
(with $3_1\#3_1$ at $3.4$ and   $3_1\#4_1$ at $7.8$).
}
\label{fig:ideal}
\end{figure*}

The relation between the typical length and the corrections to scaling may be extended
to include a further relation with the length of ideal knots.
For off-lattice knotted flexible yet impenetrable tubes of thickness $D$ and length $L$, 
the ideal length $l^{\rm id}$ of a given knot is the smallest ratio $L/D$ one 
can achieve~\cite{Katritch:1996:Nature,IdealKnotsBook}.
For a lattice knot the ``ideal'' configurations are those with the minimal number ($l^{\rm min}$) of 
steps~\cite{Diao:1993:JKTR,Janse:2011:JSM}.
The probability maxima discussed above are located at relatively small values
of $\overline{l_1}$, not too far from $l^{\rm min}_{k_1}$. This suggests that the configurations taken by
the loop with $k_1$ for $l_1\approx \overline{l_1}$ could be somehow reminiscent,
up to temperature dependent moderate deviations, of the configurations
of minimal length, with $\overline{l_1} \sim l^{\rm min}_{k_1}$. Evidence of
the plausibility of this relation is obtained by reporting our determinations of
$M_{k_1}$ divided by $(l^{\rm min}_{k_1})^{5/3}$ in Fig.~\ref{fig:ideal}(a), and
divided by $(l^{\rm id }_{k_1})^{5/3}$ in Fig.~\ref{fig:ideal}(b) (note that the
known values suggest that to a first approximation $l^{\rm min}\propto l^{\rm id}$).
Thus, the existence of the topological correction and its
dependence on topology should ultimately
follow from the fact that tight knots have a minimal length depending
primarily on $n_c$. This is but another instance in which
the ideal form of knots is in close relation with their physical behavior~\cite{IdealKnotsBook}.

We conclude by adding that in principle the terms in Eq.~(5) may be complemented
with knot-independent corrections to scaling. While we have assessed with nonlinear
fits of the distributions $P(l_1)$ that any
correction $\sim 1/l_1$ would be present with a prefactor of order unity (as opposed
the $M_k$'s, which are at least $\gtrsim 700$), there remains to understand
whether corrections such as $B/l_1^\Delta$ with $0<\Delta<1$ would play a role. Preliminary results
appear to show that $B \approx 100$ if $\Delta=1/2$ is assumed. With this magnitude, the corrections would
just slightly modify quantitatively but not qualitatively our conclusions above.

\section{Conclusions}

Globular or dense polymers are known as an hard subject in
statistical physics, from both theoretical and numerical points
of view. With the present contribution we tried to show that
they provide an unexpectedly rich context in which to pursue
the goals of topological polymer statistics.
Indeed, many of the results we obtained are rather unexpected
and surprising, in some cases also for their deceptive simplicity.
Going deeper into these issues constitutes an open challenge.

When considering rings in the globular phase, any planar projection
of the configurations yields an enormous number of crossings~\cite{Grassberger:2001:JPA},
growing like $N^{4/3}$. This huge number of crossings
makes extremely difficult the analysis
of the type of knot, which requires to sort out
the few crossings which are essential characterization of the
knot. In spite of this, we showed that the thermodynamics of
the globule is driven by the number of these minimal crossings
in several ways, and that in some situations $n_c$ can become
unexpectedly a key parameter for determining the stability of the
system.

In first place we clarified the role of topology in determining
the spectrum of globular knots in a standard model of collapsed
polymer rings. This provides a unique example so far of spectrum
in which all knots, prime or composite, enter with comparable
frequencies (for swollen rings such
balance occurs only within subclasses of knots
 with the same number of prime components).
The remarkable independence of the relative frequencies on temperature
suggests that they are  universal ratios and poses the question of their 
link with the knot topological invariants.

Another surprising result is provided by the topological free energy
correction and its dependence on the minimal crossing number $n_c$.  
Under the constraint imposed by a slip link, which does not allow trespassing of the knots from
one loop to the other, $n_c$ appears to control very sharply
and exclusively the average share of backbone between the loops.
At the same time this average behavior results from an interesting
regime of broad fluctuations obeying a form of Taylor scaling~\cite{Zoltan:2008:AP}.

The numerical experiments concerning the translocation of
the globule through a repulsive plane with a hole are perhaps
the most interesting both within the context of the present
investigation, and in the perspective of future applications.
Besides providing a strong support to the
postulated form of the topological free energy correction, our 
results indicate how far reaching can be the influence
of $n_c$ in determining behaviors such as entropic forces driven
by topology. Notably, we found that these features are related to the 
concept of ideal knots, whose signature thus emerges in the statistics
of random globular knotted polymers. 

\begin{acknowledgement}
EO acknowledges support from the Italian Ministry of Education grant PRIN 2010HXAW77.
\end{acknowledgement}


\providecommand*\mcitethebibliography{\thebibliography}
\csname @ifundefined\endcsname{endmcitethebibliography}
  {\let\endmcitethebibliography\endthebibliography}{}


\end{document}